\documentclass[10pt,twocolumn,twoside]{IEEEtran}
\ifCLASSINFOpdf
\else
\fi
\usepackage{graphicx}
\usepackage{amssymb,amsmath,latexsym,amsfonts}
\usepackage{epsfig}
\usepackage{comment}
\usepackage{color}

\newtheorem{theorem}{Theorem}
\newtheorem{lemma}{Lemma}
\newtheorem{Proposition}{Proposition}

\def\done{\hspace*{\fill} \rule{1.8mm}{2.5mm}}

\begin{document}
%
\title{Analysis of Buffer Starvation with Application to Objective QoE Optimization of Streaming Services
\thanks{Part of this work appeared in IEEE Infocom 2012. }
}
%
%
%

\author{Yuedong~Xu,
        Eitan~Altman,~\IEEEmembership{Fellow,~IEEE,}
        Rachid~El-Azouzi, Majed~Haddad, Salaheddine~Elayoubi, and~Tania~Jimenez
\IEEEcompsocitemizethanks{\IEEEcompsocthanksitem 
Yuedong Xu is with department of Electronic Engineering, Fudan University, Shanghai China. Email: ydxu@fudan.edu.cn 

Eitan Altman and Majed~Haddad are with INRIA Sophia Antipolis, France. Email: eitan.altman@inria.fr and haddadmajed@yahoo.fr

Rachid~El-Azouzi and
and Tania~Jimenez are with Universite d'Avignon, France. Email: rachid.elazouzi@univ-avignon.fr and
tania.jimenez@univ-avignon.fr

Salaheddine~Elayoubi is with Orange Labs, Paris, France. Email: salaheddine.elayoubi@orange.com}}

\maketitle

\begin{abstract}
Our purpose in this paper is to characterize buffer starvations for streaming services.
The buffer is modeled as an M/M/1 queue, plus the
consideration of bursty arrivals. When the buffer is empty,
the service restarts after
a certain amount of packets are \emph{prefetched}.
With this goal, we propose two approaches to obtain the \emph{exact distribution}
of the number of buffer starvations, one of which
is based on \emph{Ballot theorem}, and the other uses
recursive equations. 
The Ballot theorem approach gives an explicit result. We extend this approach to the 
scenario with a constant playback rate using T\`{a}kacs Ballot theorem. 
The recursive approach, though not offering an explicit result, can obtain the 
distribution of starvations with non-independent and identically distributed (i.i.d.) 
arrival process in which an ON/OFF bursty arrival process is considered in this work.
We further compute the starvation probability as a function of the amount of 
prefetched packets for a large number of files via a fluid analysis.
Among many potential applications of starvation analysis, we show how to apply it to 
optimize the objective quality
of experience (QoE) of media streaming, by exploiting the tradeoff between startup/rebuffering delay and starvations.
\end{abstract}
\noindent \textbf{EDICS:} \textbf{8-MSAT}, \textbf{3-QAUE}

%
\IEEEpeerreviewmaketitle

\section{Introduction}

The starvation probability of a buffer is an important performance measure
for protocol design of telecommunication networks, as well as in storage
systems and ecological systems (e.g. dams).
Starvation is said to occur when the buffer is empty.
Various applications use buffering in order to control the rate
at which packets are served at the destination. As long as there
are packets in the buffer, packets arrive at the destination
regularly, i.e. they are spaced by the service time of the
buffer. Once the buffer empties packets may arrive at the destination
separated by larger times, as the spacing between packets now depends
also on the inter-arrival times at the queue. Starvation is
in particular undesirable in video streaming applications.

The time till starvation of a queue is related to the busy period which
has been well studied under the assumption of a stationary arrival process
(see \cite{Baccelli, Ledermann} and their references).
In contrast to this assumption, we consider a finite number of arrivals
as we are interested in statistics of starvation when a file of fixed size
is transferred.

The main goal of this paper is to find the \emph{distribution of the
number of starvations}
within a file of $N$ packets. We first model the buffer as an M/M/1 queue, and
then extend it to incorporate the bursty packet arrival that is modeled by an \emph{interrupted
Poisson process (IPP)}. In this system, a fixed amount of packets are \emph{prefetched}
(also called ``prefetching threshold'') before the service begins or resumes after a starvation event.

In this paper, we mainly propose two approaches (that give the same result) to compute the distribution of the number of starvations for a single file. 
The first approach gives an explicit result based on the famous \emph{Ballot theorem} \cite{Takacs}. But it is in general suitable for independent and identically distributed 
(i.i.d.) arrival process. 
This motivates us to propose the second approach that yields a more flexible 
recursive algorithm.
Although it has no explicit result, it can be used to compute the starvation probabilities 
for more complicated arrival processes. In terms of complexity, 
the recursive approach can compute the starvation probabilities for different 
combinations of the initial start-up threshold and the file size in one round, which means 
an overall smaller complexity than the Ballot theorem approach.

The key feature of Ballot theorem is its simple expression to compute the 
probability that a counting process (e.g. arrival process) is strictly ahead of another 
counting process (e.g. departure process). 
Using Ballot theorem, we can compute in a simple way the exact distribution
of the number of starvations explicitly. When the file size is large enough,
we present the asymptotic starvation probability using Gaussian (interchangeable with Normal)
approximation as well as an approximation of the Riemann integral.
As a special case that the playout buffer is modeled by an M/D/1 queue, the
probabilities of the number of starvations can be obtained on the basis of a 
discrete version of Ballot theorem. 
Furthermore, unlike the Ballot theorem that usually requires the i.i.d. packet arrivals, 
the recursive approach enables us to compute
the probability of starvations with continuous-time ON/OFF bursty packet arrivals.  
The media server transfers packets during the ON state and pauses during the OFF state. 
It switches from one state to the other after residing an exponentially 
distributed time period.

We further propose a fluid analysis of starvation behavior on the file level.
This approach, instead of looking into the stochastic packet arrivals and departures,
predicts the starvation where the servers manage a large quantity of file transfers.
Given the traffic intensity and the distribution of file
size, we are able to compute the starvation probability as a function of the
prefetching threshold. The fluid analysis, though simple, offers an
important insight on how to control the probability of starvation for many
files, instead of for one particular file.

The probabilities of starvations developed in this work have various applications
in the different fields. A prominent example is the media streaming service.
This application demonstrates a dilemma between the prefetching process
and the starvation. A longer prefetching process causes a larger start-up/rebuffering delay, while
a shorter one might result in starvations. The user perceived media quality (or QoE equivalently)
is impaired by two major factors, the large start-up/rebuffering delay and the frequent starvations, according to the
measurement studies in \cite{Sigcomm11:Dobrian,IMC12:Krishnan}. These two
factors are thus defined as quality metrics of streaming service.
The QoE study becomes increasingly important in the epoch that
web video hogs up to more than 37\% of total traffic during peak hours in USA \cite{webvideo}.
In contrast to the rapid growth of traffic load, the bandwidth provision usually lags behind.
In this context, media providers and network operators face a crucial challenge of
maintaining a satisfactory QoE of streaming service.
With the results developed in this work, we are able to answer the fundamental question: \emph{
How many packets should the media player prefetch to optimize the users' quality of experience?}

To answer this question, we first use objective QoE costs to model the subjective 
human unhappiness for both finite and infinite file size. 
An objective QoE cost is the weighted sum 
functions on the start-up/rebuffering delay and the starvation behaviors. 
The weight reflects an individual user's relative impatience on the delay than on the starvations. We then formulate the objective QoE optimization problems in a variety of scenarios. 
After solving them, we obtain the optimal algorithms to 
configure the start-up/rebuffering thresholds in packets. 
Lastly, we discuss the possible implementation of the proposed algorithms. 
A streaming user can dynamically configure the rebuffering threshold. 
When a starvation happens, the packet arrival rate can be measured at the user side.
Given the knowledge the arrival rate, service rate and the remaining file size, 
the user can set the rebuffering threshold to balance the tradeoffs:  
i) delay vs starvation probability (section \ref{sec:QoE_A}) for a small remaining file size
and and ii) delay vs starvation interval (section \ref{sec:QoE_B}) for a large remaining file size. 
The media server can configure the start-up threshold for each category of 
videos (e.g. music, sports, and TV news) beforehand. 
Given the distributions of file size and the user throughput, it can use the algorithm in
section \ref{sec:QoE_C} to compute the optimal start-up threshold.
Simulation studies in section \ref{sec:QoE} demonstrate the relationships between 
the objective QoE costs and the optimal prefetching thresholds in a variety of scenarios.

We summarize the main results as follows.

\begin{itemize}

\item We propose a Ballot theorem approach to compute the distribution of the number of
starvations for a file of finite size in an M/M/1 queue. It provides an explicit solution and is 
generalized to an M/D/1 queue.

\item We propose a recursive approach to compute the distribution of the number of starvations. 
We further extend it to the ON/OFF bursty arrival process.

\item We present a fluid model to compute the starvation probability, given the start-up threshold and the file size distributions of representative content types.

\item Our study provides the important understandings on 
how the starvation 
probabilities are impacted by the prefetching threshold. We propose a set of strategies 
to configure better prefetching thresholds so as to balance the tradeoff between the 
delay and the starvations for video streaming services.

\end{itemize}

The rest of this paper is organized as follows. Section \ref{sec:related}
reviews the related work. We propose a Ballot approach in
Section \ref{sec:Ballot}. Section \ref{sec:recursive} presents the recursive
approach for an M/M/1 queue and extents it to the ON/OFF bursty arrival
process. Section \ref{sec:fluid} performs a fluid analysis for a large number of files.
Section \ref{sec:QoE} presents the QoE metrics and their optimization issues.
Our theoretical results are verified in section \ref{sec:simulation}.
Section \ref{sec:conclusion} concludes this paper.

\section{Related Work}
\label{sec:related}

The analysis of starvation is close to that
of busy period in transient queues.
In \cite{Baccelli,Ledermann} 
authors solve the distribution of the buffer size
as a function of time for the M/M/1 queue. The exact
result is expressed as an infinite sum of modified
Bessel functions.
The starvation analysis of this work is different from the
transient queueing analysis in two aspects.
First, the former aims to find the probability generating function of starvation events
while not the queue size. Second, the former does not assume
a stationary arrival process.

Ballot theorem and recursive equations have been used to analyze
the packet loss probability in a finite buffer when the
forward error-correcting technique is deployed. Citon et al.  \cite{TIT93:Citon}
propose a recursive approach that enables them to compute
the packet loss probability in a block of consecutive packet arrivals
into an M/M/1/K queue.
Based on their recursive approach, Altman and Jean-Marie in \cite{JSAC98:Eitan} obtain
the expressions for the multidimensional generating function of
the packet loss probability. The distribution of message delay
is given in an extended work \cite{Infocom95:Eitan}.
Dubea and Altman in \cite{Dubea} analyze the packet loss probability
with the consideration of random loss in incoming and outgoing links.
In \cite{Gurewitz}, Gurewitz et al. introduce the powerful Ballot theorem to find
this probability within a block of packet arrivals into an M/M/1/K queue.
They consider two cases, in which
the block size is smaller or greater than the buffer limit. Another example
of applying Ballot theorem to evaluate networking system is found in \cite{Humblet}.
Humblet et al. present a method based on Ballot theorem
to study the performance of nD/D/1 queue with periodical arrivals and deterministic
service time. In \cite{Infocom01:Sohraby}, He and Sohraby use Ballot theorem
to find the stationary probability distribution in a general class of discrete
time systems. Privalov and Sohraby \cite{PIT07:Sohraby}
study the underflow behavior of CBR traffic in a time-slotted queueing system. However,
they do not provide the insights of having a certain number of starvations.

In the applications related to our work, Stockhammer et al. \cite{TMM02:Stockhammer} specify
the minimum start-up delay and the minimum buffer size for a given video stream and
a deterministic variable bit rate (VBR) wireless channel.
Recently, \cite{TMM08:Liang} presents a deterministic bound, and
\cite{JSAC11:ParandehGheibi} provides a stochastic bound
of start-up delay to avoid starvation. Authors in \cite{TMM10:Luan} model the playout buffer
as a G/G/1 queue. By using diffusion approximation,
they obtain the closed-form starvation probability with asymptotically large file size.
Xu et.al \cite{yuedong2}
study the scheduling algorithms for multicast streaming in multicarrier wireless downlink. 
Authors in \cite{Networking12:xu} studied the QoE metrics
of a persistent video streaming in cellular networks. 
They further presented a new method in \cite{Infocom13:xu} 
to compute the 
QoE metrics for a cellular network with arrivals and 
departures of streaming flows.
In the application field, our paper differs from state of the art works in the following ways:
i) we present new theories that yield an \emph{exact} probability of starvation,
and the p.g.f. of starvation events; ii) we study
the asymptotic behavior with error analysis; iii) we perform a macroscopic starvation analysis using a fluid model; iv) we configure optimal prefetching thresholds to
optimize the QoE metrics.

\section{Starvation Analysis Using Ballot Theorem}
\label{sec:Ballot}

In this section, we study
the starvation behavior of an M/M/1 (also extended to 
M/D/1) queue with finite number of
arrivals based
on the powerful Ballot theorem.

\subsection{System Description}

We consider a single media file with finite size $N$.
The media content is pre-stored in the media server 
(e.g. video on demand (VoD) service). When a
user makes a request, the server segments this media into
packets, and transfers them to the user by use of TCP or UDP. 
When packets traverse the wired or wireless links,
their arrivals to the media player of a user are not deterministic
due to the dynamics of the available bandwidth.
The Poisson assumption is not the most realistic way to describe packet
arrivals, but it reveals the essential features of the system, and
is the first step for more general arrival processes.
After the streaming
packets are received, they are first stored in the playout buffer.
The interval between two packets that are served is assumed to be
exponentially distributed so that we can model
the receiver buffer as an M/M/1 queue.
The maximum buffer size is assumed to be large enough
so that the whole file can be stored. This simplification is justified
by the fact that the storage space is usually very large in the receiver
side (e.g. several GB).

The user perceived media quality has two measures called \emph{start-up delay}
and \emph{starvation}. As explained earlier, the media player
wants to avoid the starvation by prefetching packets.
However, this action might incur a long waiting time.
In what follows, we reveal the relationship between the start-up delay
and the starvation behavior, with the consideration of file size.

\subsection{A Packet Level Model}

We present a packet level model to investigate the starvation behavior.
We denote by $\lambda$ the Poisson
arrival rate of the packets, and by $\mu$ the Poisson service rate. We define $\rho:=\lambda/\mu$ as the traffic intensity.

In a non-empty M/M/1 queue with everlasting arrivals, the rate at which
either an arrival or a departure occurs is given by $\lambda + \mu$. This event corresponds
to an arrival with probability $p$, or is otherwise to an end of service
with probability $q$, where
\begin{eqnarray}
p = \frac{\lambda}{\lambda + \mu} = \frac{\rho}{1+\rho}; \;\;\;\;\; q = \frac{\mu}{\lambda + \mu} = \frac{1}{1+\rho}. \nonumber
\end{eqnarray}
The buffer is initially empty. We let $T_1$ be the start-up delay, in which
$x_1$ packets are accumulated in the buffer. Our analysis of the probability of 
starvation is built on the famous Ballot theorem:

\noindent \textbf{Ballot Theorem:} {\em In a ballot, candidate A scores $N_A$
votes and candidate B scores $N_B$ votes, where $N_A > N_B$. Assume
that while counting, all the ordering (i.e. all sequences of A's and B's)
are equally alike, the probability that throughout the counting, A is always
ahead in the count of votes is $\frac{N_A - N_B}{N_A + N_B}$.}

After the service begins, the probability of starvation is given
by Theorem \ref{theorem:nobuffering}.
\begin{theorem}
\label{theorem:nobuffering}
For the initial queue length $x_1$ and the total size $N$ of a file, the probability of
starvation is given by:
\begin{eqnarray}
P_{s} = \sum_{k=x_1}^{N-1} \frac{x_1}{2k-x_1}\binom{2k-x_1}{k-x_1}p^{k-x_1}(1-p)^k.
\label{eq:ballot}
\end{eqnarray}
\end{theorem}
\noindent \textbf{Proof:}
We define $E_k$ to be an event that the buffer becomes empty
for the first time when the service of packet $k$ is finished.
It is obvious that all the events $E_k, k=1,\cdots N,$ are mutually exclusive.
Then, the event of starvation is the union $\cup_{k=x_1}^{N-1} E_k$.
This union of events excludes $E_N$ because the empty buffer after the service of $N$ packets
is not a starvation. When the buffer is empty at the end
of the service of the $k^{th}$ packet, the number of arrivals is $k - x_1$
after the prefetching process. The probability of having $k-x_1$ arrivals
and $k$ departures is computed from a binomial distribution, $\binom{2k-x_1}{k-x_1}p^{k-x_1}(1-p)^k$.
We next find the necessary and sufficient condition of the event $E_k$.
If we have a backward time axis that starts from the time point when the buffer is empty for the
first time, the number of departure packets is always more than that of arrival packets.
As a result, the Ballot Theorem can be applied.
For example, among the last $m$ events (i.e. $m\leq 2k-x_1$),
the number of packets that have been played is always greater than the number of arrivals.
Otherwise, the empty buffer already happens before the $k^{th}$ packet is served.
According to the Ballot theorem, the probability of event $E_k$ is computed by
$\frac{x_1}{2k-x_1}\binom{2k-x_1}{k-x_1}p^{k-x_1}q^k$. Therefore, the probability
of starvation, $P_{s}$, is the probability of the union $\cup_{k=x_1}^{N-1} E_k$, given by
eq.\eqref{eq:ballot}.
\done

\begin{figure*}[!htb]
    
    \includegraphics[width=5in]{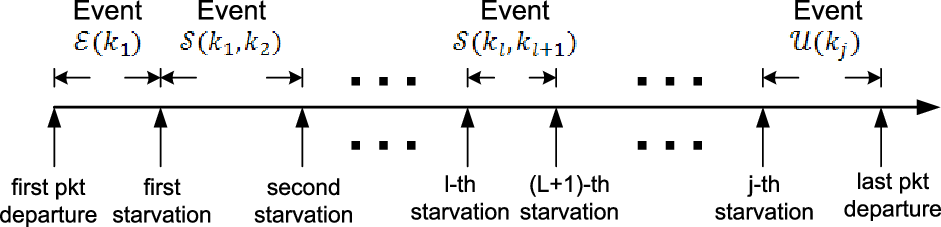}
    \centering
    \caption{A path with $j$ starvations}
    \label{fig:samplepath}
    \vspace{-0.3cm}
\end{figure*}

The starvation event may happen for more than once during the file transfer. We are particularly
interested in the probability distribution of starvations, given a finite file size $N$. The maximum
number of starvations is $J=\lfloor\frac{N}{x_1}\rfloor$ where $\lfloor\cdot\rfloor$
is the floor of a real number. We define \emph{path} as a complete sequence of
packet arrivals and departures. The probability of a path depends
on the number of starvations. We illustrate a typical path with $j$ starvations in Figure \ref{fig:samplepath}.
To carry out the analysis, we start from the event that the first starvation takes place. We denote
by $k_l$ the $l^{th}$ departure of a packet that sees an empty queue. We notice that the path
can be decomposed into three types of mutually exclusive events as follows:
\begin{itemize}

\item Event $\mathcal{E}(k_1)$: the buffer becoming empty for the first time in the entire path.

\item Event $\mathcal{S}_l(k_l,k_{l+1})$: the empty buffer after the service of packet $k_{l+1}$ given
that the previous empty buffer happens at the departure of packet $k_{l}$.

\item Event $\mathcal{U}_j(k_j)$: the last empty buffer observed after the departure of packet $k_j$.

\end{itemize}
Obviously, a path with $j$ starvations is composed of a succession of events
$$\mathcal{E}(k_1), \mathcal{S}_1(k_1, k_2), \mathcal{S}_2(k_2, k_3), \cdots, $$
$$\mathcal{S}_{j-2}(k_{j-2}, k_{j-1}), \mathcal{S}_{j-1}(k_{j-1}, k_{j}), \mathcal{U}_j(k_j).$$
\noindent We let $P_{\mathcal{E}(k_1)}$, $P_{\mathcal{S}_l(k_l, k_{l+1})}$ and $P_{\mathcal{U}_j(k_j)}$ be
the probabilities of events $\mathcal{E}(k_1)$, $\mathcal{S}_l(k_l, k_{l+1})$ and $\mathcal{U}_j(k_j)$
respectively. The main difficulty to analyze the probability mass function is that the media player
pauses for $x_1$ packets upon starvation.
In what follows, we analyze the probabilities of these events step by step.
The event $\mathcal{E}(k_1)$ can happen
after the departure of packet $k_1 \in [x_1, N-1]$. According to the proof of Theorem \ref{theorem:nobuffering},
the probability distribution of event $\mathcal{E}(k_1)$ can be expressed as
\begin{eqnarray}
P_{\mathcal{E}(k_1)} := \left\{\begin{matrix}
0  &&\!\!\!\!\!\!\!\!\!\!\!\!\!\!\!\!\!\!\!\!\!\!\!\!\!\!\textrm{ if } k_1 < x_1 \textrm{ or  } k_1 = N ;\\
\frac{x_1}{2k_1-x_1}\binom{2k_1-x_1}{k_1-x_1}p^{k_1-x_1}q^{k_1}  &&\textrm{ otherwise }.
\end{matrix}\right.
\label{eq:probeventE}
\end{eqnarray}
The first starvation cannot happen at the departure of first $(x_1{-}1)$ packets because of the prefetching of $x_1$ packets. It cannot happen after all $N$ packets have been served because this empty buffer is not a starvation.
We next solve the probability distribution of the event
$\mathcal{U}_j(k_j)$. Suppose that there are $j$ starvations after the service of packet $k_j$.
The extreme case is that these $j$ starvations take place consecutively. Thus,
$k_j$ should be greater than $jx_1-1$. Otherwise there cannot have $j$ starvations. 
The starvation event cannot take place after the departure of packet 
$N$ because the whole file has now been transferred.
If $k_j$ is no less than $N-x_1$, the media player resumes until all the remaining 
$N-k_j$ packets are stored
in the buffer. Then, starvation will not appear afterwards. In the remaining cases,
the event $\mathcal{U}_j(k_j)$ is equivalent to the event that no starvation happens after the
service of packet $k_j$. We can take the complement of starvation
probability as the probability of no starvation. Hence, the probability distribution of event
$\mathcal{U}_j(k_j)$ is given by
\begin{eqnarray}
P_{\mathcal{U}_j(k_j)} := \left\{\begin{matrix}
\!\!\!0,  \;\;\;\;\;\;\textrm{ if } k_j < jx_1   \textrm{ or } k_j=N;\\
1,  \;\;\;\;\textrm{ if } N-x_1 \leq k_j < N ;\\
1 - \sum^{N-k_j-1}_{m=x_1}\frac{x_1}{2m-x_1}\binom{2m-x_1}{m}p^{m-x_1}q^m, \\
\textrm{ otherwise }.
\end{matrix}\right.
\label{eq:probeventU}
\end{eqnarray}
\noindent We denote by $P_s(j)$ the probability of having $j$ starvations. The probability
$P_s(0)$ can be obtained from Theorem \ref{theorem:nobuffering} directly. For the case
with one starvation, $P_s(1)$ is solved by
\begin{eqnarray}
P_s(1) = \sum^{N}_{i=1} P_{\mathcal{E}(i)} P_{\mathcal{U}_1(i)}  = \mathbf{P}_{\mathcal{E}} \cdot \mathbf{P}_{\mathcal{U}_1}^{T}
\label{eq:onestarvation}
\end{eqnarray}
\noindent where $^T$ denote the transpose. 
Here, $\mathbf{P}_{\mathcal{E}}$ is the row vector of $P_{\mathcal{E}(i)}$,
and $\mathbf{P}_{\mathcal{U}_1}$ is the row vector of $P_{\mathcal{U}_1(i)}$,  for $i=1,2,\cdots, N$.

To compute the probability of having more than one starvations, we need to find the probability of
event $\mathcal{S}_l(k_l,k_{l+1})$ beforehand.
Solving $P_{\mathcal{S}_l(k_l, k_{l+1})}$
is non-trivial due to that the probability of this event depends on the remaining file size
and the number of starvations. After packet $k_l$ is served, the $l^{th}$ starvation is observed.
It is clear that $k_l$ should not be less than $lx_1$ in order to have $l$ starvations.
Given that the buffer is empty after serving packet $k_l$,
the $(l+1)^{th}$ starvation cannot happen at $k_{l+1} \in [k_l+1, k_l+x_1-1]$ because of
the subsequent prefetching process. Since there are $j$ starvations
in total, the $(l+1)^{th}$ starvation must satisfy $k_{l+1} <  N-(j-l-1)x_1$.
We next compute the remaining case that the $l^{th}$ and the $(l+1)^{th}$ starvations happen
after packets $k_l$ and $k_{l+1}$ are served. Then, there are $(k_{l+1}-k_l)$ departures, and
$(k_{l+1}-k_l-x_1)$ arrivals after the prefetching process. According to the Ballot theorem, a path without starvation
between the departure of packet $(k_l+1)$ and that of packet $(k_{l+1})$ is
expressed as $\frac{x_1}{2k_{l+1}-2k_l-x_1}$. Therefore, we can express $P_{\mathcal{S}_l(k_l, k_{l+1})}$ as
\begin{eqnarray}
\left\{\begin{matrix}
\frac{x_1}{2k_{l+1}-2k_l-x_1}\binom{2k_{l+1}-2k_l-x_1}{k_{l+1}-k_l-x_1} p^{k_{l+1}-k_l-x_1}q^{k_{l+1}-k_l}, \\
\;\;\textrm{ if } k_l \geq lx_1, k_l+x_1 \leq k_{l+1} < N-(j-l-1)x_1 ;\\
0, \;\;\;\;\;\;\;\;\textrm{ otherwise }.
\end{matrix}\right.
\label{eq:probeventS}
\end{eqnarray}

\noindent We denote by $\mathbf{P}_{\mathcal{S}_l}$ the matrix of $P_{\mathcal{S}_l(k_l, k_{l+1})}$ for $k_l, k_{l+1} \in [1,N]$.
Here, $\mathbf{P}_{\mathcal{S}_l}$ is an upper triangle matrix where all the elements in the first $(lx_1-1)$ rows, and the last $x_1$ rows are 0.
The probability of having $j (j\geq 2)$ starvations is given by
{\small
\begin{eqnarray}
P_s(j) \!\!\!&=&\!\!\! \sum^{N}_{k_1=1} \sum_{k_2=1}^{N}\cdots \sum_{k_{j-1}=1}^{N}\sum_{k_j=1}^{N} P_{\mathcal{E}(k_1)}\cdot P_{\mathcal{S}_1(k_1, k_{2})} \cdots  \nonumber\\
\!\!\!&&\!\!\! P_{\mathcal{S}_{j-1}(k_{j-1}, k_{j})}\cdot  P_{\mathcal{U}_j(k_j)}
= \mathbf{P}_{\mathcal{E}}\Big(\prod_{l=1}^{j-1} \mathbf{P}_{\mathcal{S}_l}\Big) \mathbf{P}_{\mathcal{U}_j}^{T}.
\label{eq:jstarvation}
\end{eqnarray}
}
The probability of no starvation, $P_s(0)$, is computed as $1{-}P_s$ where 
$P_s$ is obtained from eq.\eqref{eq:ballot}. 
Since the starvation event takes non-negative integer values, we can write the probability generating function (p.g.f.) $G(z)$ by
\begin{eqnarray}
G(z) = E(z^j) = \sum\nolimits_{j=0}^{J} P_s(j) \cdot z^j .
\label{eq:generatingfunc}
\end{eqnarray}
\noindent In $\mathbf{P}, \mathbf{P}_{\mathcal{S}_l}$ and $\mathbf{P}_{\mathcal{U}_j}$,
the binomial distributions can be approximated by the corresponding Normal distributions
with negligible errors (see Appendix).
The Gaussian approximation significantly reduces
the computational complexity of binomial distributions.

We next analyze the complexity of matrix (including vector) operations in eq. 
\eqref{eq:jstarvation}. 
A matrix operation consists of floating-point operations where 
one floating-point operation can be an addition, subtraction, multiplication or division of 
two float type matrix elements \cite{book_matrix}. In the complexity analysis of matrix operations, the 
lower-order terms are usually ignored. The approximated probability of
starvation in eq.\eqref{eq:ballot} consists of $N$ additions, 
thus having a complexity order $O(N)$.
The probability of one starvation is a product of two vectors, which consists of $N$ 
multiplications and $N{-}1$ additions. Hence, the complexity order in still $O(N)$.
If there are only two starvations, we need to compute
the product of two vectors and one matrix, which has a complexity order $O(N^2)$.
When $j\geq 3$, the computation of
$P_s(j)$ involves the product of two matrices. In general, multiplying two matrices
has a complexity order $O(N^{3})$. In eq.\eqref{eq:jstarvation}, we should multiply 
a vector and a matrix each time, instead of multiplying two matrices inside the bracket.
Then, the matrix operations in eq.\eqref{eq:jstarvation} only contain a set of multiplications
between a vector and a matrix. This yields the complexity order $O(N^2)$. To sum up, 
the computation of the p.g.f. of starvations in eq.\eqref{eq:generatingfunc} has a 
complexity order $O(N^2)$, given the start-up/rebuffering threshold $x_1$ 
and the file size $N$.

\noindent \textbf{Asymptotic Property:}

We want to know whether the starvation event yields simple implications
as the file size $N$ approaches $\infty$. The
asymptotic behavior of the starvation probability is given by
\begin{eqnarray}
\lim_{N\rightarrow \infty}P_{s} := \left\{\begin{matrix}
1 \;\;\; &&\textrm{ if } \rho < 1 ;\\
\exp\big(\frac{x_1(1-2p)}{2pq}\big) \;\;\; &&\textrm{ otherwise }.
\end{matrix}\right.
\label{eq:probstarv_asymp}
\end{eqnarray}
\noindent The detailed analysis can be found in the Appendix.

The asymptotic starvation probability is irrelevant to 
the start-up threshold when $\rho < 1$.
Under this situation, it is necessary to know
how frequent the starvation event happens. Here, we compute
the average time interval between two starvations. We let $T_s$ be the
duration of starvation interval. Its expectation $E[T_s]$ is the
expected busy period of an M/M/1 queue with $x_1$ customers in the beginning \cite{JAP96:Liu}, i.e.
\begin{eqnarray}
E[T_s] = \frac{x_1}{\lambda(1-\rho)}.
\end{eqnarray}

\subsection{Extension to Discrete-time Systems}

In general, the playback rate of video streaming has a much smaller
variance than the arrival rate. Hence, the playback of streaming
packets is sometimes regarded as a time-slotted process (e.g. \cite{TMM08:Liang,JSAC11:ParandehGheibi}) 
where only one packet is served at the beginning of a time slot.
We consider a playout buffer modeled as an M/D/1 queue.
We denote by $d$ the duration of a slot.

In this subsection, we introduce a discrete Ballot theorem named
Tak\'{a}cs Ballot Theorem.
\begin{theorem} (Tak\'{a}cs Ballot Theorem \cite{Takacs})
\label{theorem:takacs}
If $X(1), X(2), \cdots$, $X(l)$ are cyclically interchangeable r.v.s taking on nonnegative integer
values summing to $k$, then
\begin{eqnarray}
\mathbb{P}\Big\{\sum_{s=1}^{t}X(s) < t, \forall \; t\in [1, l]\Big\} = \frac{[l-k]^+}{l}. 
\end{eqnarray}
\end{theorem}
The Tak\'{a}cs Ballot Theorem presents a probability that the number of departures
is larger than that of arrivals in all $l$ slots.
If the arrival process $\{X(s)\}$ is Poisson, $X(s)$ is i.i.d. at different slots, and thus cyclically interchangeable.
Suppose that the starvation event happens after $t$ packets have been served, ($t\geq x_1$). 
The total number of arrivals is $t{-}x_1$.
We create a backward time axis where the starvation event happens at slot 1. The number 
of departures is always greater than that of arrivals. Otherwise, the starvation event 
has already taken place. Hence, according to Tak\'{a}cs Ballot theorem, the probability of the departure always
leading the arrival  is
\begin{eqnarray}
\mathbb{P}\big\{\sum_{s=1}^{t}X(s) < t, \; \forall t\in [1, l]\big\} = \frac{x_1}{l}.
\end{eqnarray}
Therefore, the probability that the first starvation takes place after the service of 
the $l^{th}$ packet (i.e. starvaton event happening at slot $(l{+}1)$)
\begin{eqnarray}
P_s(l) = \frac{x_1}{l} \cdot \mathbb{P}\{\sum_{s=1}^{l} X(s)= l{-}x_1\}, \quad \forall l\geq x_1 .
\label{eq:takacs_starvprob}
\end{eqnarray}
For the Poisson process $\{X(s)\}$, the probability of $l{-}x_1$ 
packet arrivals in $l$ slots (i.e. the duration $ld$) is obtained by
\begin{eqnarray}
\mathbb{P}\{\sum_{s=1}^{l} X(s)= l{-}x_1\} = \frac{(\lambda ld)^{l{-}x_1} }{(l{-}x_1)!}\exp(-\lambda ld).
\label{eq:takacs_poisson}
\end{eqnarray}
Given the file size $N$ and the prefetching threshold $x_1$, the starvation might happens upon the departure of packets from $x_1$ to $N-1$.
Then the starvation probability is obtained by
\begin{eqnarray}
P_{s} = \sum_{l=x_1}^{N-1} \frac{x_1}{l}\frac{(\lambda ld)^{l{-}x_1} }{(l{-}x_1)!}\exp(-\lambda ld).
\label{eq:takacs_ballot}
\end{eqnarray}

We next show how the p.g.f. of starvation events can be derived using the Tak\'{a}cs Ballot theorem.
The path with $j$ starvations is the same as that in Fig.\ref{fig:samplepath}. With certain abuse of notations,
we reuse $\mathcal{E}(k_1)$, $\mathcal{U}_j(k_j)$ and $\mathcal{S}_l(k_l,k_{l+1})$ to denote
the first, the last and the other starvation events. 
According to eq.\eqref{eq:takacs_poisson}, there has 
\begin{eqnarray}
P_{\mathcal{E}(k_1)} :=
 \left\{\begin{matrix}
0  &&\!\!\!\!\!\!\!\!\!\!\!\!\!\!\!\!\!\!\!\!\!\!\!\!\!\!\!\!\!\!\textrm{ if } k_1 < x_1 \textrm{ or  } k_1 = N ;\\
 \frac{x_1}{k_1}\frac{(\lambda k_1d)^{k_1{-}x_1} }{(k_1{-}x_1)!}\exp(-\lambda k_1d)   &&\textrm{ otherwise }.
\end{matrix}\right.
\label{eq:probeventE_cbr}
\end{eqnarray}
Since there exist $j$ starvations in total, the last starvation event will not happen at the departure of packets less than $jx_1$. Given the last starvation happening as soon as the $k_j^{th}$ packet is served, the probability of no
starvation afterwards can also be solved using eq.\eqref{eq:takacs_poisson}.
\begin{eqnarray}
P_{\mathcal{U}_j(k_j)} := \left\{\begin{matrix}
\!\!\!0,  \;\;\;\;\;\;\textrm{ if } k_j < jx_1   \textrm{ or } k_j=N;\\
1,  \;\;\;\;\textrm{ if } N-x_1 \leq k_j < N ;\\
1{-} \sum^{N{-}k_j{-}1}_{s=x_1}\frac{x_1}{s}\frac{(\lambda sd)^{s{-}x_1} }{(s{-}x_1)!}\exp({-}\lambda sd), \\
\textrm{ otherwise }.
\end{matrix}\right. \nonumber
\label{eq:probeventU_cbr}
\end{eqnarray}
When the $l^{th}$ and the $l{+}1^{th}$ starvation events appear at the departure of 
packet $k_l$ and $k_{l{+}1}$, the probability $P_{\mathcal{S}_l(k_l, k_{l+1})}$ is given by
\begin{eqnarray}
\left\{\begin{matrix}
\frac{x_1}{k_{l{+}1}-k_{l}}\frac{(\lambda (k_{l{+}1}-k_{l})d)^{(k_{l{+}1}-k_{l}{-}x_1)}}{(k_{l{+}1}-k_{l}{-}x_1)!}
\exp({-}\lambda (k_{l{+}1}{-}k_{l})d)\\
\;\;\textrm{ if }k_l \geq lx_1, k_l+x_1 \leq k_{l+1} < N-(j-l-1)x_1 ;\\
0, \;\;\;\;\;\;\;\;\textrm{ otherwise }.
\end{matrix}\right.
\label{eq:probeventS_cbr}
\end{eqnarray}
Then, the p.g.f. of starvation events can be solved using eq.\eqref{eq:generatingfunc} in the same way.

\section{Starvation Analysis Via a Recursive Approach}
\label{sec:recursive}

In this section, we present a recursive approach to 
compute the p.g.f. of starvations based on
\cite{TIT93:Citon}. 
Compared with the one using Ballot theorem,
the recursive approach can handle more complicated 
arrival process.

\subsection{Probability of Starvation}

The probability of starvation and the p.g.f can be analyzed all in once.
However, we compute them separately because the analysis of
the starvation probability provides an easier route to understand
this approach.

We denote by $P_i(n)$ the probability of starvation with a file of $n$ packets,
given that there are $i$ packets in the system just before the arrival epoch
of the first packet of this file. In the original system, our purpose
is to obtain the starvation probability of a file with the size $N$ when
$x_1$ packets are prefetched before the service begins. This corresponds to
$P_i(n)$ with $n=N{+}1{-}x_1$ and $i=x_1-1$. \emph{Here, the expression $i=x_1-1$
means that the service starts when the $x_1$-th packet sees $x_1-1$ packets accumulated
in the buffer. When the service begins, there are already $x_1$ packets in the queue.} 
To compute $P_i(n)$, we will introduce recursive equations.
We define a quantity $Q_i(k)$, $i=0,1,\cdots, n$, $0\leq k\leq i$, which is the probability
that $k$ packets out of $i$ leave the system during an inter-arrival period.
This probability is equivalent to the probability of
$k$ Poisson arrivals with rate $\mu$ during an exponentially distributed period
with parameter $1/\lambda$. According to \cite{Book:Papoulis}, we obtain
\begin{eqnarray}
Q_i(k) &=& \rho \big(\frac{1}{1+\rho}\big)^{k+1} = pq^k , \;\; 0\leq k \leq i-1, \\
Q_i(i) &=& \big(\frac{1}{1+\rho}\big)^{i} = q^i.
\end{eqnarray}

To carry out the recursive calculation, we start from the case $n=1$.
\begin{eqnarray}
P_i(1) = 0, \;\;\; \forall i \geq 1.
\end{eqnarray}
When the file size is 1 and the only packet observes a non-empty queue, the probability
of starvation is 0 obviously. If $i$ is 0, the starvation happens for sure, thus yielding
\begin{eqnarray}
P_0(n) = 1, \;\;\; \forall n.
\end{eqnarray}
\noindent For $n\geq 2$, we have the following recursive equations:
\begin{eqnarray}
P_i(n) = \sum_{k=0}^{i+1}Q_{i+1}(k) P_{i+1-k}(n-1), \;\;\;0\leq i \leq N-1.
\label{eq:recursive1}
\end{eqnarray}
\noindent We explain \eqref{eq:recursive1} as the following. When the first packet
of the file arrives and sees $i$ packets in the system, the starvation does not happen.
However, the starvation might happen in the service of remaining $n-1$ packets.
Upon the arrival of the next packet, $k$ packets out of $i+1$ leave the system
with probability $Q_{i+1}(k)$. We next add constraints to the recursive equation \eqref{eq:recursive1}
for a file of size $N$. Since the total number of packets is $N$, the
starvation probability must satisfy $P_i(n) = 0$ for $i+n > N$.

\subsection{P.G.F. of Starvations}

To compute the p.g.f. of starvation, we use the same recursive approach, despite of
the more complicated structure. With certain reuse of notation, we
denote by $P_i(j,n)$ the probability of $j$ starvation of a file with size $n$, given
that the first packet of the file sees $i$ packets in the system upon its arrival.
Our final purpose is to compute the probability of starvation for a file of size $N$.
It can be obtained from $P_i(j,n)$ with $i=x_1{-}1$ and $n=N{+}1{-}x_1$. 
When the first packet of remaining $n$ packets arrives at the buffer,
it sees $x_1{-}1$ packets. At this time point, there are $x_1$ packets in the buffer and 
the service of packets begins. 

In order to compute $P_i(j,n)$ recursively, we provide the initial conditions first:
\begin{eqnarray}
P_{i}(j,1) = \left\{\begin{matrix}
0 \; &&\forall i=1,2,\cdots, N-1, \textrm{ and } j\geq 1;\\
1 \; &&\forall i=1,2,\cdots, N-1, \textrm{ and } j = 0,
\end{matrix}\right.
\label{eq:recursive2}
\end{eqnarray}
\noindent and
\begin{eqnarray}
P_{0}(j,1) = \left\{\begin{matrix}
0 \; && j = 0 \textrm{ or } j\geq 2;\\
1 \; && j =1.
\end{matrix}\right.
\label{eq:recursive3}
\end{eqnarray}
\noindent The equation \eqref{eq:recursive2} means that
the probability of no starvation is 1 conditioned by $i\geq 1$ and $n=1$.
Thus, the probability of having one or more starvations is 0 obviously if
the only packet sees a nonempty system. The equation \eqref{eq:recursive3}
reflects that the starvation happens for sure when the only packet
observes an empty queue. However, there can only have one starvation event
due to $n=1$. Another practical constraint is
\begin{eqnarray}
P_{i}(j,n) = 0, \;\;\; \textrm{ if } i+n > N
\label{eq:recursive4}
\end{eqnarray}
\noindent because of the finite file size $N$.

To compute $P_{i}(j,n)$, we need to know what will happen if
the buffer is empty, i.e. $i=0$. One intuitive observation is
\begin{eqnarray}
P_{0}(0,n) = 0, \;\;\; \forall \;1\leq n\leq N-b;
\label{eq:recursive5}
\end{eqnarray}
\noindent where $b:=x_1-1$ is denoted to be the prefetching threshold.
Eq.\eqref{eq:recursive5} holds because an empty queue means at least one starvation event.
For a more general probability $P_{0}(j,n)$, we begin with the case
$j=1$. If $n\leq b$ and the first packet of $n$ sees an empty buffer, there has
only one starvation, that is,
\begin{eqnarray}
P_{0}(1,n) = 1, \;\;\; \forall \;1\leq n\leq b,
\label{eq:recursive5}
\end{eqnarray}
If $n>b$, $b$ packets will be prefetched. Thus, the remaining file
size is $n-b$. We see $b$ packets in the system upon the arrival
of the first packet in the remaining file.
Given that the only one starvation
event has taken place, there will be no future starvations. Therefore,
the following equality holds,
\begin{eqnarray}
P_{0}(1,n) = P_{b}(0,n-b), \;\;\; \forall \;b<n\leq N-b.
\label{eq:recursive6}
\end{eqnarray}
\noindent Using the similar method, we can solve $P_{0}(j,n)$ for $j>1$.
However, the property of $P_{0}(j,n)$ with $j>1$ is quite different
\begin{eqnarray}
P_{0}(j,n) = 0, \;\;\; \forall \; j>1 \textrm{ and } 1\leq n\leq b.
\label{eq:recursive7}
\end{eqnarray}
This means that the probability of having $>1$ starvations is 0 if
the file size is no larger than $b$. If $n$ is greater than $b$,
then $b$ packets are prefetched, leaving $n-b$ packets in the remaining file.
The remaining $n-b$ packets encounter $j-1$ starvations, given that the
first packet sees $b$ packets in the system upon arrival, i.e.
\begin{eqnarray}
P_{0}(j,n) = P_{b}(j-1,n-b), \;\;\; \forall \; j>1 \textrm{ and } n > b.
\label{eq:recursive8}
\end{eqnarray}

So far, we have computed a critical quantity $P_0(j,n)$, the probability of meeting
an empty buffer. Next, we construct recursive equations to compute  $P_i(j,n)$
as the following:
\begin{eqnarray}
\!\!\!\!\!\!\!\!\!\!\!\!&&P_i(j,n) = \sum_{k=0}^{i+1} Q_{i+1}(k) P_{i+1-k}(j, n-1), \nonumber\\
\!\!\!\!\!\!\!\!&&=\sum_{k=0}^{i} pq^k P_{i+1-k}(j, n-1) + q^{i+1}P_{0}(j, n-1),
\label{eq:recursive9}
\end{eqnarray}
\noindent for $0\leq i \leq N-1$.
The eq.\eqref{eq:recursive9} contains two parts. The former expression
reflects the cases that the next arrival sees an \emph{non-empty} queue.
The latter one characterizes the transition of the system to a prefetching
process that is computed by \eqref{eq:recursive8}.

We are interested in how efficient the recursive method is. Hence, we present
the roadmap to compute $P_i(j,n)$ and its complexity:
\begin{itemize}
\item \textbf{Step 1:} Solving $P_i(0,2)$, for $i=1$ to $N-2$;
\item \textbf{Step 2:} Solving $P_i(0,n)$, for $i=1$ to $N-2$, and $n=3$ to $N{-}x_1{+}1$ based on \emph{Step 1};
\item \textbf{Step 3:} Adding $j$ by 1 and computing $P_i(j,n)$ based on \emph{Step 1} and \emph{Step 2}.
\end{itemize}
The complexity analysis is carried out from this roadmap. 
An operation in this recursive algorithm refers to an addition. In \textbf{step 1}, 
the computation of $P_i(0,2)$ incurs $i{+}2$ additions for each $i$ according to 
eq.\eqref{eq:recursive9}. Hence, the total number of additions for all $i$ from 1 to 
$N$ is around $N^2/2$.
\textbf{Step 2} computes $P_i(0,n)$ repeatedly for each $n\in\{2,\cdots,N\}$.  The \textbf{Step 3} repeats \textbf{Step 1\&2} for each $j$, but not augmenting the 
complexity order in $N$. Therefore, the total complexity has the order $O(N^3)$. 
The recursive algorithm obtains the starvation probabilities 
for all $j$ and $n$, ($j\leq J$, $n\leq N$)  and all initial start-up threshold 
$i$ ($1\leq i \leq N$).

\noindent \textbf{Remark 1:}  We compare the complexity of the Ballot approach and the recursive approach. First, the standard Ballot approach 
contains factorial terms (e.g. $\binom{2k-x_1}{k-x_1}$ in eq.\eqref{eq:ballot}) 
that are of high computational burdens. Second, 
after Gaussian approximation of factorial terms, the Ballot approach has a complexity 
order $O(N^2)$ given the file size and the start-up threshold. 
The recursive approach has a complexity $O(N^3)$ for all combinations of initial start-up
threshold $i$ and file size $n (1\leq i,n\leq N)$. Thus, the recursive approach has 
an overall smaller complexity than the Ballot  approach.

\subsection{ON/OFF Bursty Traffic}

In this section, we model the arrival process as an \emph{interrupted Poisson process (IPP)},
which is commonly used to characterize the bursty and correlated arrivals. The source
may stay for relatively long durations in ON and OFF states. The ON/OFF arrival model
also has direct applications. For example, the Youtube servers use a simple ON/OFF rate control
algorithm to transfer streaming packets to the users \cite{CCR11:Alcock}. Our objective is to understand
the interaction between the parameters of arrival process and the probability of starvation.

\begin{figure}[!htb]
    \centering
    \includegraphics[width=2.5in]{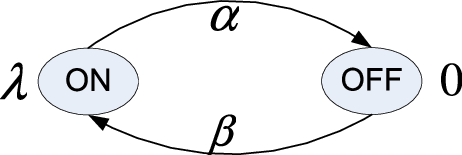}
    \caption{Two-state Markov process to model bursty traffic}
    \label{fig:onoff}
\end{figure}

We illustrate the bursty traffic model in figure \ref{fig:onoff} with the state transition
rates $\alpha$ and $\beta$. 
We denote by $Q_i(k)^{ON}$, $0\leq i\leq N-1$, $0\leq k\leq i$, the probability that $k$ packets
out of $i$ leave the system upon an arrival at the ON state (i.e. no arrival during the OFF period).
According to \cite{TIT93:Citon}, the following proposition holds.
\begin{Proposition}\cite{TIT93:Citon}
\label{Proposition:no1}
The probability $Q_i(k)^{ON}$ is expressed as
\begin{eqnarray}
Q_i(k)^{\mathrm{ON}} &=& c_1\big(\frac{1}{a_1}\big)^k  + c_2\big(\frac{1}{a_2}\big)^k, \;\;\; 0\leq k \leq i-1, \nonumber\\
Q_i(i)^{\mathrm{ON}} &=& c_1\frac{(1/a_1)^i}{1-1/a_1} + c_2\frac{(1/a_2)^i}{1-1/a_2},
\end{eqnarray}
\noindent where $a_1$, $a_2$, $c_1$ and $c_2$ are solved by
\begin{eqnarray}
\Delta &=& (\lambda + \alpha + \beta)^2 - 4\lambda\beta, \nonumber\\
a_{1,2} &=& 1+ \frac{\lambda+\alpha+\beta}{2\mu} \pm \frac{\sqrt{\Delta}}{2\mu}, \nonumber\\
c_1 &=& \frac{\lambda(\beta+\mu)-\lambda\mu a_1}{a_1(a_2-a_1)}, \;\;c_2 = \frac{\lambda(\beta+\mu)-\lambda\mu a_2}{a_2(a_1-a_2)}.\nonumber
\end{eqnarray}
\end{Proposition}

We next show how the starvation probability $P_i(j,n)$ is obtained. The starvation
event can happen in both the ON and OFF states.
However, the starvation event at the OFF state is equivalent to the event that
the first new packet arrival at the ON state sees an empty queue. Therefore, we can use
\eqref{eq:recursive7} to compute the p.g.f. of starvations with bursty arrivals,
simply replacing $Q_i(k)$ by $Q_i^{\mathrm{ON}}(k)$.

\noindent \textbf{Remark 2:} The standard Ballot theorem cannot
be used to study the starvations of the ON/OFF arrival process. 
In the presence of bursty traffic, the packet arrival process has two states, 
ON and OFF. The packet arrivals of two consecutive ON states are separated by an OFF state.
Hence, the counting of the arrival arrivals is not equally probable.

\section{Fluid Model Analysis of Starvation Probability}
\label{sec:fluid}

So far we have studied the starvation behavior of a single file, which is concerned
by either the media servers or the users. In fact, the streaming providers are more
interested in the QoE evaluation scaled to a large quantity of videos.
They cannot afford the effort of configuring each file a different start-up delay.
In this section, we present a fluid analysis of starvation probability, given the
distribution of file size.

In the fluid model, the arrival and departure rates are deterministic.
We let $\lambda$ be the number of packet
arrivals \emph{per second}, and $\mu$ be the number of departures \emph{per second}.
Here, $\mu$ depends on the encoding rate that the media files use.
We focus on the setting $\mu \geq \lambda$ because no starvation will happen
with $\mu < \lambda$ in the fluid model. We let $x_1$ be the start-up threshold.
The start-up delay $T_1$ is simply computed by $x_1/\lambda$. 
Once the media packets are played, the queue length decreases at a rate $\mu - \lambda$.
The time needed to empty the queue is thus $\frac{x_1}{\mu-\lambda}$. We let $N_p$
be the total number of packets that are served until a starvation happens,
\begin{eqnarray}
N_p = x_1\big( 1+ \frac{\lambda}{\mu-\lambda}\big ) = \frac{x_1 \mu}{\mu - \lambda}.
\end{eqnarray}
\noindent If the file size is less than $N_p$, there will be no starvation event.

The distribution of media file size depends on the types of contents. A measurement study
in \cite{IWQoS08:Cheng} shows the distributions of Youtube video duration for four most
popular categories: music, entertainment, comedy and sports videos.
The authors find that most of the entertainment, comedy and sports videos are short.
They are likely to follow exponential distribution or lognormal distribution with large standard
deviations. The lengths of music files in playback time are usually between 180 and 240 seconds
(the file size in Bytes is the product of duration and the default bit-rate on Youtube).
Hence, with the help of the measurements in \cite{IWQoS08:Cheng}, we speculate that music video
files on Youtube follow lognormal distribution with a small standard deviation. The Pareto distribution 
adds practical restrictions to the video file size. The file size needs to be greater than a 
certain value, and a very small fraction of video files can be very large. 
Today, Youtube allows some users to upload some long movies (more than the previous maximum of 10 min) without copyright issues. 
The distribution of movie file size may have a heavy-tail. Then,
Pareto distribution can serve as a good approximation.

We compare the starvation probabilities of exponential, log-normal and Pareto distributions, given the start-up threshold.
Note that these distributions possess the same mean file size. We assume
that the users are homogeneous so that $\lambda$ and $\mu$ are the same for
different types of file size distributions.

i) \emph{Exponential distribution:} Suppose that the file size $N$
follows an exponential distribution with parameter $\theta$. The probability
of starvation, $P_s^{(1)}$, is obtained by
\begin{eqnarray}
P_s^{(1)} = \textrm{Prob }(N > N_p) = \exp(- \frac{\theta x_1 \mu}{\mu - \lambda}).
\label{eq:prob_starv_exp}
\end{eqnarray}

ii) \emph{Pareto distribution:} We let $N_m$ be the minimum
possible value of the file size, and $\upsilon$ be the exponent in the
Pareto distribution. The probability of starvation is computed by
\begin{eqnarray}
P_s^{(2)} = \textrm{Prob }(N > N_p) = \left\{\begin{matrix}
\big(\frac{N_m(\mu-\lambda)}{\mu x_1}\big)^{\upsilon} \!\!\! && \forall N_m \leq \frac{x_1 \mu}{\mu - \lambda};\\
1  \!\!\!&& \textrm{ otherwise },
\end{matrix}\right.
\label{eq:prob_starv_pareto}
\end{eqnarray}
\noindent where the expectation of the Pareto distribution
is equal to that of the exponential distribution, i.e. $\frac{\upsilon N_m}{\upsilon-1} = \frac{1}{\theta}$.

iii) \emph{Log-Normal distribution:} We suppose that the file size
follows a log-normal distribution $\ln \mathcal{N}(\varrho,\sigma)$,
where $\varrho$ and $\sigma$ are the mean and the standard deviation
of a natural normal distribution. Given that $N_p$ packets can be served without an interruption,
the starvation probability $P_{s}^{(3)}$ is computed by
\begin{eqnarray}
P_s^{(3)} = \textrm{Prob }(N > N_p) = \frac{1}{2} - \frac{1}{2}\mathrm{erf}\big[\frac{\log \frac{x_1 \mu}{\mu - \lambda} -\varrho}{\sqrt{2}\sigma}\big],
\label{eq:prob_starv_lognormal}
\end{eqnarray}
\noindent where its expectation $\exp(\varrho+\frac{\sigma^2}{2})$ equals to $\frac{1}{\theta}$.

Equations \eqref{eq:prob_starv_exp},\eqref{eq:prob_starv_pareto} and
\eqref{eq:prob_starv_lognormal} show that the probability of starvation can be
controled  by setting $x_1$,
if the distribution of file size, the arrival and departure rates
are pre-knowledge\footnote{Because the starvation probabilities $P_s^{(1)}$, $P_s^{(2)}$ and $P_s^{(3)}$ take complicated forms, we
will compare their dependency on $x_1$ numerically in section \ref{sec:simulation}.
Both Pareto and Log-normal distributions have two parameters. In the comparison, we fix
one of them, and solve the other according to the property of identical expectations.}.

\section{Application to Streaming Service}
\label{sec:QoE}

This section presents four scenarios in streaming service in which
our analysis can be utilized to optimize the objective QoE. Here, we focus on the M/M/1 system.

The QoE reflects the human perception of the streaming service. A common
practice to evaluate QoE is called Mean Opinion Score (MOS). 
The video watchers give scores according to their subjective opinions. 
The start-up delay and the starvation behaviors are explicitly defined
as quality metrics related to user perception 
in \cite{Sigcomm11:Dobrian}, \cite{IMC12:Krishnan} and \cite{JSAC11:ParandehGheibi}.
To remove the confusion, we designate the direct human perception as 
subjective QoE, and designate the objective measure as objective QoE.
The human tests implicitly map the objective QoE metrics into a single subjective value.
This QoE value, though revealing the user perception statistically,
is usually unreliable to report the QoE for each individual watching.
At the same time, the subjective test is done after the watching, which cannot be
utilized to tune the prefetching online. Therefore, a rising trend is to evaluate 
the objective QoE metrics and to balance the tradeoff among them (e.g. 
\cite{Sigcomm11:Dobrian}, \cite{IMC12:Krishnan} and \cite{JSAC11:ParandehGheibi}).

Our purpose is to use content prefetching as a way to achieve the optimal tradeoff 
between the start-up delay and the starvation behaviors (either the starvation probability or the
continuous playback interval) for a user. 
In \cite{JSAC11:ParandehGheibi}, the authors configure
a start-up threshold to guarantee that the starvation probability
is less than a certain value. The bound of the starvation probability is deemed 
as a parameter obtained from human tests. In this paper, we adopt a more flexible method by defining
an objective QoE cost function for a user. A user-defined weight $\lambda$ is introduced to indicate his/her preference
to one type of objective QoE metrics. We first let the starvation probability
be one of the QoE metrics.
We let $g(\cdot)$ be a strictly increasing but convex function of the expected start-up delay $E[T_1]$.
The larger the start-up delay, the higher the QoE cost. The convexity of $g(\cdot)$ means 
that streaming users are more and more impatient to large start-up delays. 
We denote by $C_1(x_1)$ the cost of a user watching the media stream,
\begin{eqnarray}
C_1(x_1) = P_s  + \gamma g(E(T_1)),
\label{eq:QoE_metric1}
\end{eqnarray}
\noindent where $\gamma$ is a positive constant. A large $\gamma$ represents that
the users are more sensitive to the start-up delay, and a smaller $\gamma$ means
a higher sensitivity to the starvation. Our goal is to find the optimal start-up threshold
$x_1^*$ to minimize $C_1(x_1)$. 

The choice of $C_1(x_1)$ should satisfy three basic principles. First, it is convex in $x_1$ so that only one
optimal threshold $x_1^*$ exists. Second, $C_1(x_1)$ is bounded even if $\rho$ is close to 1. Otherwise,
the configuration of $x_1$ is extremely sensitive to $\rho$. Third, though $x_1^*$ is not required to be
a decreasing function of the arrival rate $\lambda$, it cannot grow unbounded when $\lambda$ is large enough.
In what follows, we simply let
$g(E(T_1)):= (E(T_1))^2 = \big(\frac{x_1}{\lambda}\big)^2$.
Note that building a completely convincible cost function is very difficult for a particular user. Even the 
measurement studies in \cite{Sigcomm11:Dobrian} and \cite{IMC12:Krishnan} only quantify the influence of $P_s$ and $E(T_1)$ 
statistically and indirectly (using user engagement). We choose $C_1(x_1)$ to be the sum
of the measures of two types of objective QoE metrics for two reasons. One is the reflection of QoE tradeoff. The other
is its simplicity. Other forms of $C_1(x_1)$ can be optimized in the same way. 
The more realistic cost function is subject to our future study.

We apply our models to optimize objective QoE in three scenarios: i) finite media streaming, ii) everlasting
media streaming and iii) file level. The scenarios i) and ii) are designed for a single stream, while
iii) is designed for a large number of streams. When the streaming file has a finite size, the congested
bottlenecks such as the 3G base station or the wifi access point can configure or suggest a start-up threshold before
the media stream is played. If the steaming file is large enough (e.g. realtime sport channel), a user can 
measure the arrival/service processes, and then configure the rebuffering threshold. In the third scenario,
the media server can set up one same start-up threshold for all the videos that it distributes.
To avoid malfunctions in realistic scenarios, a user can configure the maximum and the minimum start-up/rebuffering delay. Once maximum value is reached, the media player starts to play regardless
of the prefetching threshold.

\subsection{Finite Media Size}
\label{sec:QoE_A}
We hereby consider the adaptive buffering technique for a streaming with finite size.
The eq.\eqref{eq:ballot} and eq.\eqref{eq:QoE_metric1} yield
{\small
\begin{eqnarray}
C_1(x_1) = \sum_{k=x_1}^{N-1} \frac{x_1}{2k-x_1}\binom{2k-x_1}{k-x_1}p^{k-x_1}(1-p)^k + \gamma (\frac{x_1}{\lambda})^2.
\label{eq:totalcost_finite}
\end{eqnarray}}
The starvation probability decreases and the start-up delay increases strictly as $x_1$ grows.
In the objective QoE optimization of finite media size, there does not exist a simple expression of the optimal
threshold $x_1^*$. To find $x_1^*$ numerically, we need to compare the costs from all possible thresholds.
The complexity order is low if the binomial distribution in eq.\eqref{eq:ballot}
is replaced by the Gaussian distribution. If a user can tolerate up to $1$ starvations,
$P_s$ will be replaced by the probability $(P_s(0)+P_s(1))$ according to eq.\eqref{eq:onestarvation}.

\subsection{Infinite Media Size}
\label{sec:QoE_B}
We revisit the user perceived streaming quality
in two scenarios: 1) $\rho \geq 1$ and 2) $\rho < 1$.

\noindent \textbf{Case 1: $\rho \geq 1$.} The starvation probability converges
to a fixed value when the file size approaches infinity.
We adopt the same QoE metric as that of the finite media size.
Note that $P_s$ can be directly replaced by its asymptotic value in eq.\eqref{eq:probstarv_asymp}.
Submitting $P_s$ to $C_{1}(x_1)$, we have the following cost function
\begin{eqnarray}
C_{1}(x_1) = \exp\big(\frac{x_1(1-2p)}{2pq}\big) + \gamma (\frac{x_1}{\lambda})^2.\nonumber
\end{eqnarray}
\noindent Letting the derivative $\frac{dC_{1}}{dx_1}$ be 0, we obtain
\begin{eqnarray}
x_1\cdot\exp\big(\frac{x_1(2p-1)}{2pq}\big) = \frac{(2p-1)\lambda^2}{4\gamma pq}. \nonumber
\end{eqnarray}
\noindent The optimal threshold $x_1^*$ is solved by
\begin{eqnarray}
x_1^* = LambertW\big((\frac{(2p-1)\lambda}{2 pq})^2\cdot \frac{1}{2\gamma}\big)\cdot\frac{2pq}{2p-1},\nonumber
\end{eqnarray}
\noindent where $LambertW(\cdot)$ is the Lambert W-function.

\noindent \textbf{Case 2: $\rho < 1$.} When $\rho < 1$, $P_s$
is 1 for an infinite media size. If we adopt the QoE metric $C_{1}$ directly,
the optimal start-up delay is always 0. This requires a new objective QoE
metric for the case $\rho < 1$. Since the starvation
happens many times, the continuous playback interval can serve as a measure of users' satisfaction.
We denote by $C_2(x_1)$ the cost function for an infinite media size with $\rho < 1$,
\begin{eqnarray}
C_2(x_1) := \exp(-\frac{\delta x_1}{\lambda(1-\rho)}) + \gamma (\frac{x_1}{\lambda})^2,\nonumber
\end{eqnarray}
\noindent where $\delta$ is a user defined weighting factor to the expected playback duration
($\delta:=1$ in our numerical examples).
We differentiate $C_2(x_1)$ over $x_1$, and let the derivative be 0, then the optimal start-up/rebuffering threshold is
\begin{eqnarray}
x_1^* = LambertW\big( \frac{\delta^2}{2\gamma(1-\rho)^2}\big)\cdot\frac{\lambda(1-\rho)}{\delta}.\nonumber
\end{eqnarray}

\subsection{Optimal Objective QoE in the File Level}
\label{sec:QoE_C}
Unlike the above QoE optimizations, the threshold $x_1$ for many files is configured
by the media server, instead of the users. The objective is still to
balance the tradeoff between the start-up delay and the starvation probability.
Here, only the exponentially distributed file size is considered.
We choose the cost function $C_{1}(x_1)$ that yields $C_{1}(x_1) = \exp(- \frac{\theta x_1 \mu}{\mu - \lambda})
+ \gamma \big(\frac{x_1}{\lambda}\big)^2$. The optimal threshold $x_1^*$ can be easily found as
\begin{eqnarray}
x_1^* = LambertW\big( (\frac{\theta\mu\lambda}{\mu-\lambda})^2\cdot\frac{1}{2\gamma}\big)\cdot\frac{\mu-\lambda}{\mu\theta}.\nonumber
\end{eqnarray}

\section{Numerical Examples}
\label{sec:simulation}

\subsection{Starvation of M/M/1 Queue}

\begin{figure}[!htb]
    \centering
   \includegraphics[width=2.7in, height = 2.0in]{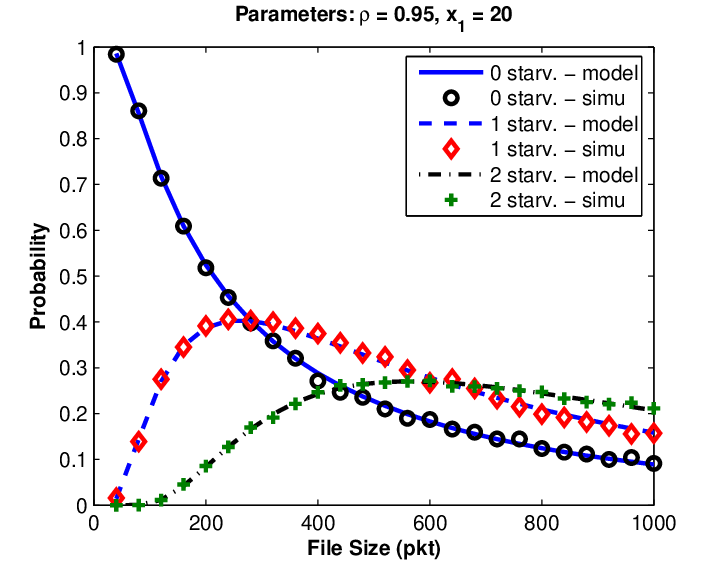}
   \caption{Probability of 0, 1, and 2 starvations with $\rho = 0.95$ and $x_1 = 20$}
   \label{fig:prob_rho095b20}
\end{figure}

\begin{figure}[!htb]
    \centering
   \includegraphics[width=2.7in, height = 2.0in]{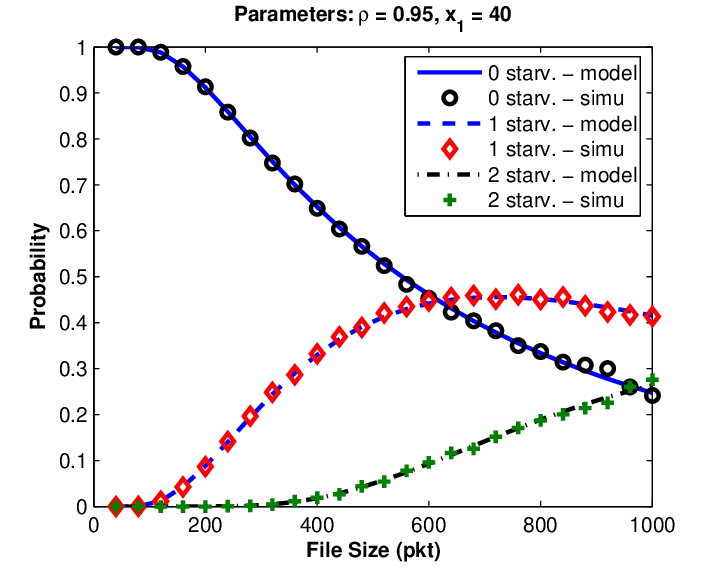}
   \caption{Probability of 0, 1, and 2 starvations with $\rho = 0.95$ and $x_1 = 40$}
   \label{fig:prob_rho095b40}
\end{figure}

This set of experiments compares the probability of starvations with the
event driven simulations using MATLAB. The M/M/1 queue is tested for up to 5000 times
with arrivals from files of different sizes. We deliberately consider four combinations
of parameters: $\rho = 0.95$ or $1.1$, and $x_1 = 20$ or $40$ pkts.
The departure rate $\mu$ is normalized as 1 if not mentioned explicitly. The choice
of the start-up thresholds coincides with the playout of audio or video streaming services
in roughly a couple of seconds (e.g. 200$\sim$400kbps playback rate on average given
the packet size of 1460 bytes in TCP).
The file size in the experiments ranges between 40 and 1000 in terms of packets.
Figure \ref{fig:prob_rho095b20} displays the probability of 0,1, and 2 starvations
with parameters $\rho = 0.95$ and $x_1 = 20$. When the file size grows, the probability
of no starvation decreases. We observe that
the probabilities of 1 and 2 starvations increase first, and then decline after
reaching the maximum values. The reason lies in that the traffic intensity
$\rho$ is less than 1. Figure \ref{fig:prob_rho095b20} also shows that our analytical
results match the simulation well. Figure \ref{fig:prob_rho095b40} exhibits the similar
results when the start-up threshold is 40 pkts. The comparison between figure
\ref{fig:prob_rho095b20} and \ref{fig:prob_rho095b40} manifests that
a larger $x_1$ is very effective in reducing starvation probability.

\begin{figure}[!htb]
   \centering
   \includegraphics[width=2.7in, height = 2.0in]{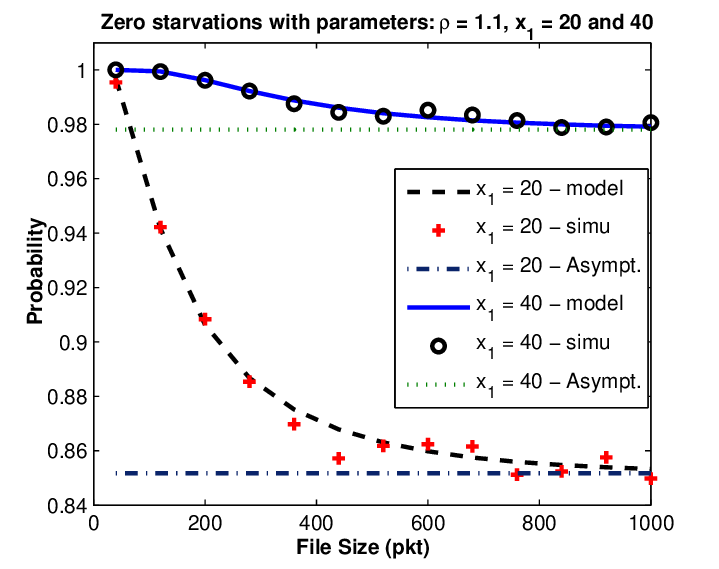}
   \caption{Probability of no starvation with $\rho = 1.1$: $x_1 = 20$ and $x_1 = 40$}
   \label{fig:prob_rho110nostarv}
\end{figure}

\begin{figure}[!htb]
    \centering
   \includegraphics[width=2.7in, height = 2.0in]{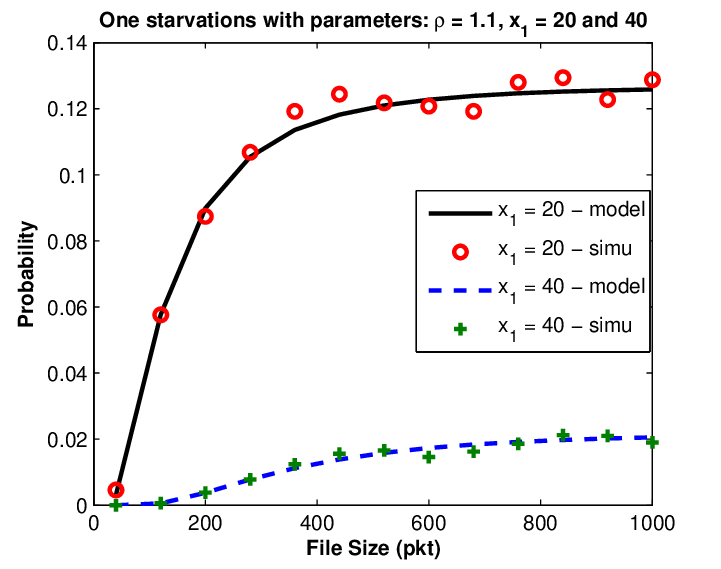}
   \caption{Probability of one starvation with $\rho = 1.1$: $x_1 = 20$ and $x_1 = 40$}
   \label{fig:prob_rho110onestarv}
\end{figure}

Figure \ref{fig:prob_rho110nostarv} plots the probability of no starvation
with the traffic intensity $\rho=1.1$. The probability of no
starvation is improved by more than 10\% (e.g. $N\geq 300$)
when $x_1$ increases from 20 to 40. Figure \ref{fig:prob_rho110nostarv}
also validates the asymptotic probability of no starvation
obtained from Gaussian and Riemann integral approximations etc.
Figure \ref{fig:prob_rho110onestarv} plots the probability of one
starvation with the same parameters. Recall that the probability
of one starvation decreases to 0 as $N$ increases in the case $\rho=0.95$.
While figure \ref{fig:prob_rho110onestarv} exhibits a different trend
along with the increase of file size. This probability becomes saturated, instead of
decreasing to 0. When $\rho$ is greater than 1, the probability of
having a particular number of starvations approaches a constant.
In both figure \ref{fig:prob_rho110nostarv} and \ref{fig:prob_rho110onestarv},
simulation results validate the correctness of our analysis. Hence, in the following experiments,
we only illustrate the analytical results.

\begin{figure}[!htb]
    \centering
   \includegraphics[width=2.7in, height = 2.0in]{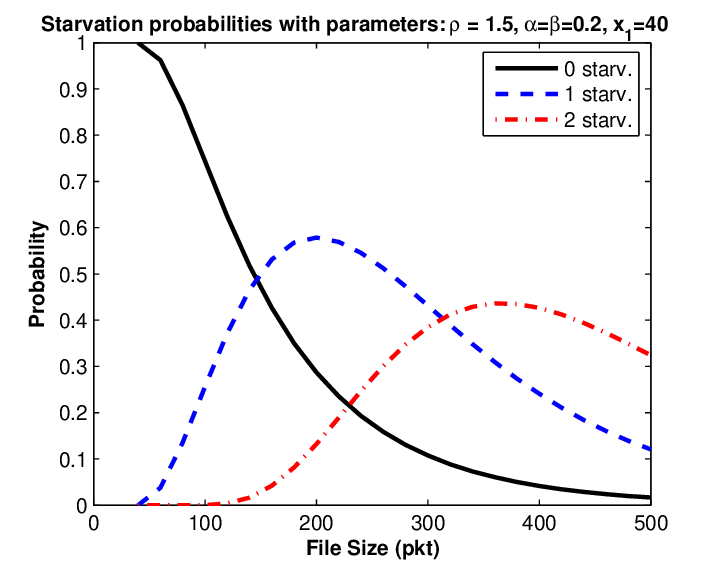}
   \caption{ON/OFF traffic: probability of 0, 1, and 2 starvations with $\rho=1.5$ and $x_1=40$}
   \label{fig:prob_onoff}
\end{figure}

\begin{figure}[!htb]
   \centering
   \includegraphics[width=2.7in, height = 2.0in]{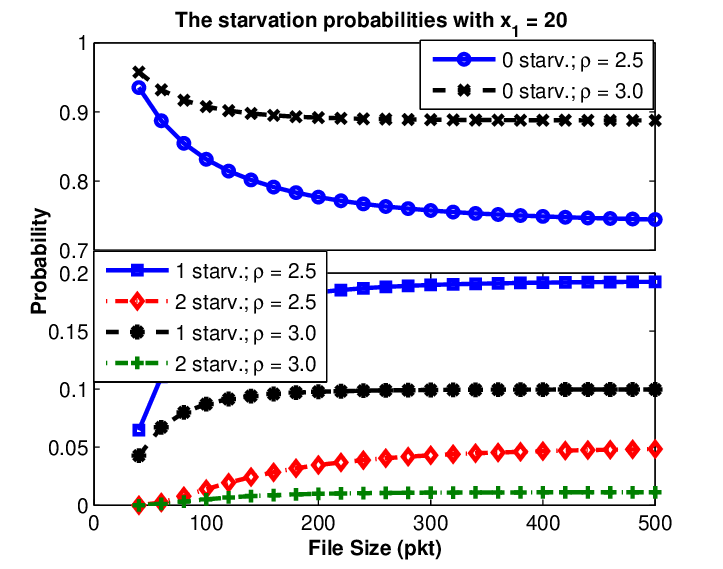}
   \caption{ON/OFF traffic: prob. of 0, 1, and 2 starvations for $x_1=20$: $\rho=2.5$ and $3.0$}
   \label{fig:prob_onoff2}
\end{figure}

\begin{figure}[!htb]
   \centering
   \includegraphics[width=2.7in, height = 2.0in]{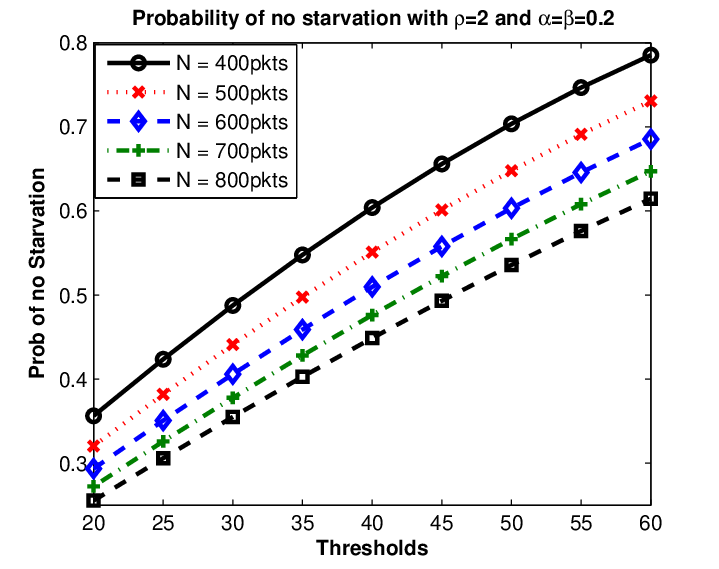}
   \caption{ON/OFF traffic: probability of no starvation with $\rho=2$ versus the threshold $x_1$}
   \label{fig:onoff_thresholdvsnostarv}
\end{figure}

\begin{figure}[!htb]
    \centering
   \includegraphics[width=2.7in, height = 2.0in]{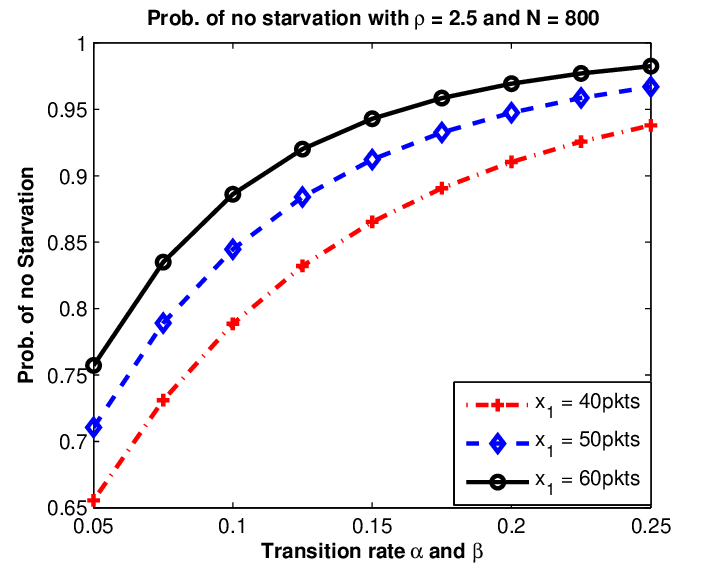}
   \caption{ON/OFF traffic: probability of no starvation with $\rho = 2.5$ and $N=800$ versus the state transition rates}
   \label{fig:onoff_abvsnostarv}
\end{figure}

\begin{figure}[!htb]
    \centering
   \includegraphics[width=2.7in, height = 2.0in]{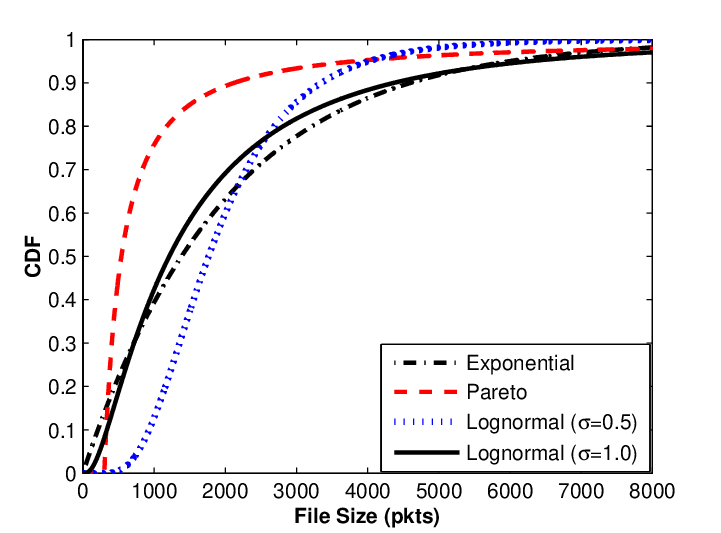}
   \caption{Fluid analysis: CDF of media file size}
   \label{fig:fluidfilesize}
\end{figure}

\begin{figure}[!htb]
    \centering
   \includegraphics[width=2.7in, height = 2.0in]{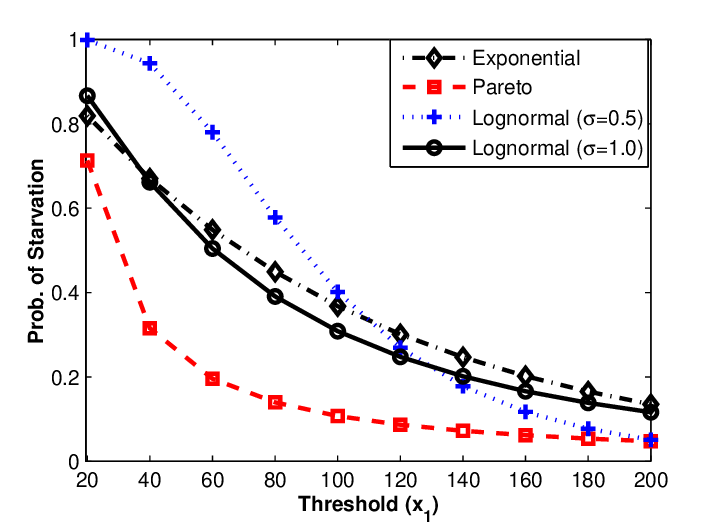}
   \caption{Fluid analysis: prob. of starvation versus the threshold $x_1$}
   \label{fig:fluidstarv}
\end{figure}

\subsection{Starvation of Bursty Traffic}

We consider the ON/OFF bursty arrival of packets into an M/M/1 queue.
For the ease of comparison, we let the transition rates $\alpha$
and $\beta$ be both 0.2. The file size ranges from 40 to 500 pkts.
In figure \ref{fig:prob_onoff}, we plot
the probabilities of having no more than two starvations with $\rho = 1.5$
and $x_1=40$. As the file size increases, the probability of no starvation decreases.
The probabilities of 1 and 2 starvations increases first, and then decreases to 0.
This means that the starvation is for sure when the file size approaches infinity.
In figure \ref{fig:prob_onoff2}, we plot the starvation probabilities for $\rho = 2.5$ and $3.0$
where the start-up threshold is set to 20. In contrast to figure \ref{fig:prob_onoff},
the probability of no starvation converges to a positive constant as $N$ is large enough.

Figure \ref{fig:onoff_thresholdvsnostarv} illustrates the impact of $x_1$ on the probability
of no starvation. In this set of experiments, $\rho$ is set to 2. The start-up threshold
$x_1$ increases from 20 to 60 pkts, and the file size increases from 400 to 800 pkts.
It is clearly shown that a slight increase in $x_1$ can greatly improve the starvation probability.
In figure \ref{fig:onoff_abvsnostarv}, we plot the probability of no starvation with
$\rho = 2.5$ and $N=800$ pkts. The transition rates $\alpha$ and $\beta$ increases from
0.05 to 0.25. It can be seen that the probability of no starvation increases monotonically
with the symmetric transition rates $\alpha$ and $\beta$.

\subsection{Starvation in the File Level}

This set of numerical experiments shows the relationship between the starvation probability
and the distribution of file size. The traffic intensity $\rho$ is set to 0.95.
We let $\theta$ be 1/2000 in the exponential distribution.
Then, the average file size is 2000 pkts. This setting is in accord with the recent
measurement that most of mobile streaming files are short (i.e. a median size of 
1.68MBytes) \cite{Infocom12:Liu}. For the Pareto distribution, we set the minimum
file size to be 300 pkts so that the exponent $\upsilon$ is 1.1765. The parameters
of the Log-normal distribution are set to $(\varrho,\sigma)= (7.476,0.5)$ and $(7.101,1.0)$ respectively. We plot the CDF curves of the file sizes in Figure 
\ref{fig:fluidfilesize}. In this setting, the Pareto distribution exhibits an obvious heavy tail 
property. For the log-normal distribution, more than 30\% percent of files are less than 
1000 packets with $\sigma = 1$ and most of the files are less than 2000 packets with 
$\sigma = 0.5$. The exponential distribution 
and the log-normal distribution with $\sigma=1.0$ exhibit similar CDF of file sizes. 
We evaluation the starvation probabilities in Figure \ref{fig:fluidstarv} 
by increasing the start-up threshold from 20 to 200. The starvation probability of the 
Pareto distributed files has a sharp reduction in the beginning of increasing $x_1$. However, as $x_1$ 
is more than 80, there is only a slight improvement in the starvation probability. Hence, 
the streaming service providers need to configure different start-up 
thresholds for the short files and the tail files in the Pareto distribution. 
For the log-normal distribution with $\sigma=0.5$, the starvation probability is
high with small $x_1$. There have significant reductions of starvation probability when 
$x_1$ increases from 40 to 140. This is because the file sizes have a small standard deviation.
Very small thresholds do not help to reduce the starvation probability and very large 
thresholds do not further reduce the starvation probability. When $x_1$ increases from 20 to 200, we 
can observe the noticeable reduction of starvation probability in the exponential file size distribution.
As the take-home message of fluid analysis, the choice of $x_1$ relies
on the distribution of file size to a great extent. To obtain a better objective QoE, the media service providers
can set different $x_1$ for different categories of media files.

\subsection{Optimizing Quality of Experience}
\label{sec:qoe_optm}
\noindent \textbf{Objective QoE optimization of finite media size:}

\begin{figure}[htb]
    \centering
   \includegraphics[width=2.7in, height = 2.0in]{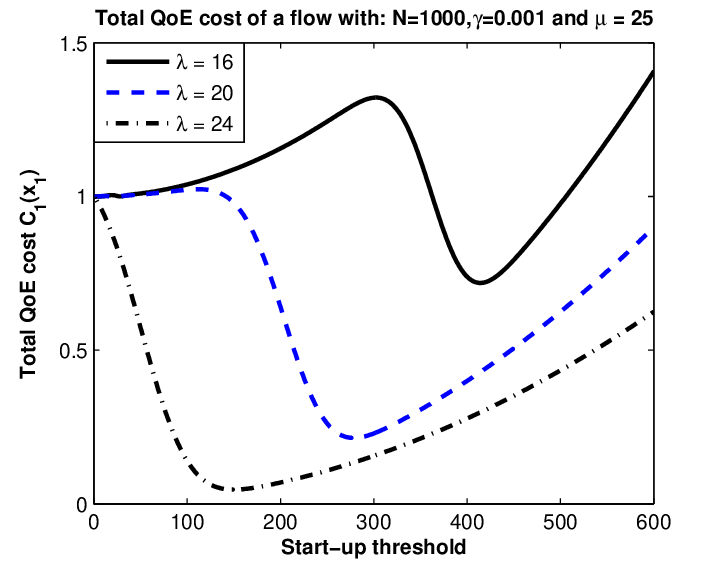}
   \caption{Finite media size: total cost with $\mu = 25$ and $\gamma = 0.001$}
   \label{fig:finite_qoe1}
\end{figure}

\begin{figure}[htb]
    \centering
   \includegraphics[width=2.7in, height = 2.0in]{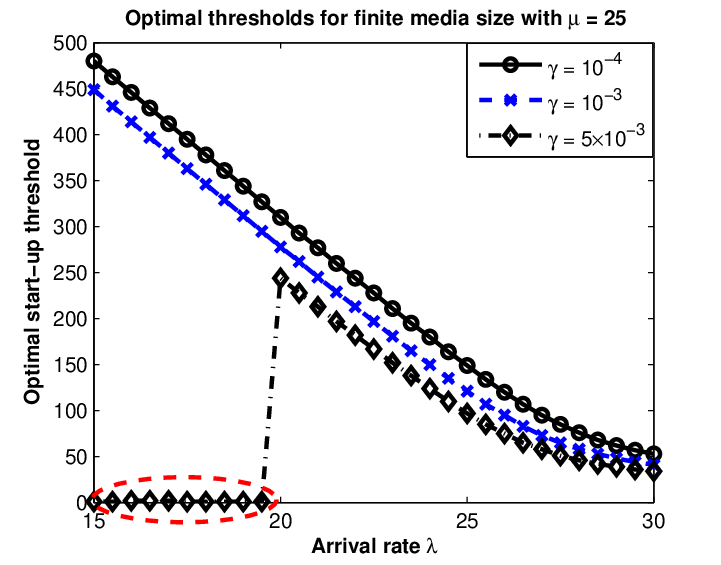}
   \caption{Finite media size: optimal thresholds with $\gamma = 10^{-4}, 10^{-3}$, and $5\times10^{-3}$}
   \label{fig:finite_qoe2}
\end{figure}

We illustrate the total QoE cost (including the starvation cost and the start-up delay cost)
in figure \ref{fig:finite_qoe1} with $\lambda = 16,20,24$
and $\mu=25$. The file size is set to $N = 1000$ and the weight $\gamma$ is $10^{-3}$.
We find that the total QoE looks neither ``concave'' nor ``convex'' w.r.t. the start-up threshold.
For example, when $x_1$ is less than 300 with $\lambda = 16$, the increase in
the start-up delay cost cannot be compensated by the reduction of the starvation probability.
We further plot the optimal start-up threshold obtained from the maximum of eq.\eqref{eq:totalcost_finite} in figure
\ref{fig:finite_qoe2}. When $\gamma = 10^{-4}$ and $\gamma = 10^{-3}$,
the optimal start-up threshold $x_1^*$ decreases when $\lambda$ increases.
We also observe that for each $\lambda$ $x_1^*$ of the case $\gamma = 10^{-4}$ is higher than that of $\gamma = 10^{-3}$
because the former user is more sensitive to the starvation. In the extreme scenario
$\gamma = 0$, the streaming user will download the whole media file before watching it.
In the case $\gamma = 5\times10^{-3}$,
$x_1^*$ are always 1 if the arrival rate $\lambda$ is less than 20. The reason lies in
that the total cost is always greater than 1 in those situations. The start-up threshold
$x_1^* = 1$ will induce numerous consecutive starvations, which definitely degrades the
streaming QoE. To mitigate this malfunction, we can introduce a minimum playback delay
that works independent of the QoE optimization.

\noindent \textbf{QoE optimization of infinite media size:}

\begin{figure}[!htb]
    \centering
   \includegraphics[width=2.7in, height = 2.0in]{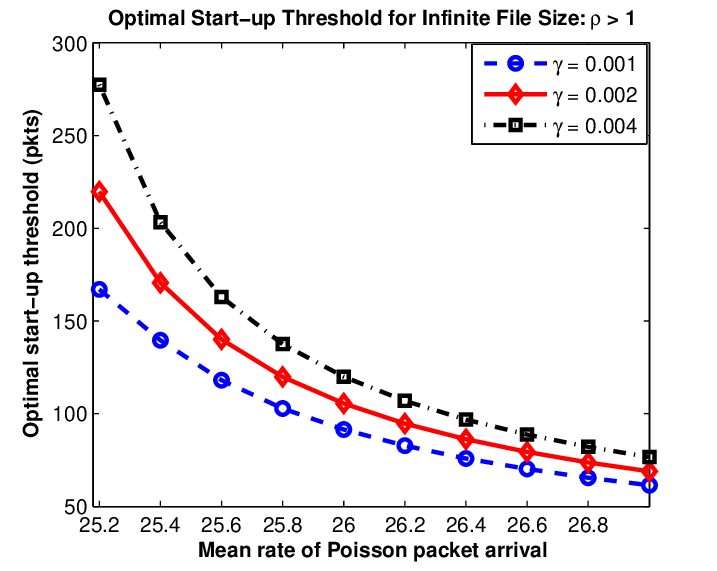}
   \caption{Optimal threshold $x_1^*$ of infinite file size: $\rho > 1$}
   \label{fig:inf_qoe_threshold1}
\end{figure}

\begin{figure}[!htb]
    \centering
   \includegraphics[width=2.7in, height = 2.0in]{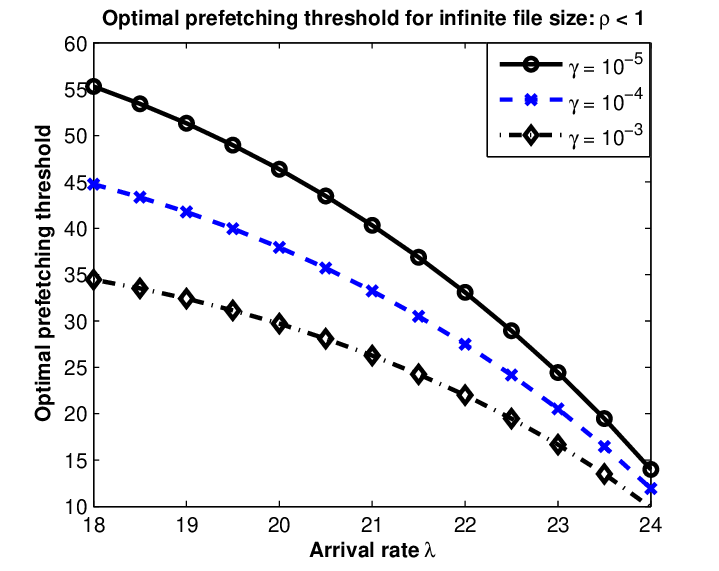}
   \caption{Optimal threshold $x_1^*$ of infinite file size: $\rho < 1$}
   \label{fig:inf_qoe_threshold3}
\end{figure}

We plot the optimal prefetching thresholds $x_1$ for the case $\rho>1$ in figure \ref{fig:inf_qoe_threshold1}
and the case $\rho<1$ in figure \ref{fig:inf_qoe_threshold3}. As $\lambda$ increases,
the optimal prefetching threshold $x_1^*$ reduces. Unlike figure \ref{fig:finite_qoe2},
there do not exist an abrupt change in $x_1^*$. This is because the cost function $C_1(x_1)$
is a convex function of $x_1$ with both $\rho>1$ and $\rho<1$.
Furthermore, $x_1^*$ decreases as $\gamma$ increases (the user putting more weight to
the prefetching delay).

\noindent \textbf{QoE optimization in the flow level:}

\begin{figure}[!htb]
    \centering
   \includegraphics[width=2.7in, height = 2.0in]{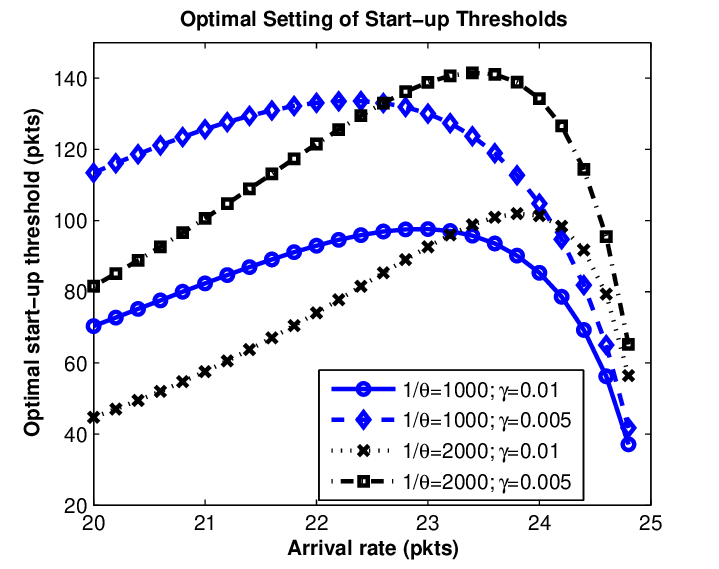}
   \caption{Optimal threshold $x_1^*$ for QoE enhancement at the file level: $\mu=25$}
   \label{fig:fluid_qoe_threshold}
\end{figure}

\begin{figure}[!htb]
    \centering
   \includegraphics[width=2.7in, height = 2.0in]{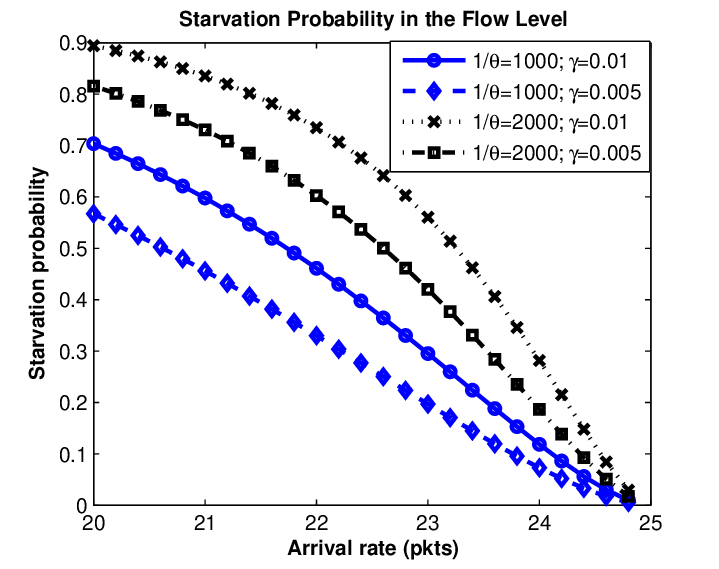}
   \caption{Starvation probability at the file level for the optimal start-up threshold $x_1^*$: $\mu=25$}
   \label{fig:fluid_qoe_starvprob}
\end{figure}

We investigate the cost minimization problem at the media server side numerically.
We let $\mu:=25$ which means that 25 packets are served per second. Given the packet size
of 1460 bytes, this service rate is equivalent to 292Kbps (without considering protocol overheads).
We let the mean file size $1/\theta$ be 1000 and 2000 packets respectively (equivalent
to the playback time of 40 and 80 seconds). The sensitivity $\gamma$ is set to 0.01 or
0.005. Figure \ref{fig:fluid_qoe_threshold} illustrates the choice of the optimal start-up thresholds
when $\lambda$ increases from 20 to 25 (i.e. $\rho\leq 1$). We evaluate four combinations of $\theta$ and $\gamma$ numerically.
Our observations are summarized as follows.
First, for the same file size distribution, a smaller $\gamma$ causes a higher optimal start-up threshold.
Second, $x_1^*$ is not a strictly decreasing function of $\lambda$. When $\lambda$ is small (e.g. 20pkts/s),
a large start-up threshold does not help much in reducing the starvation probability, but causes
impatience of users of waiting the end of prefetching. If $\lambda$ increases, the adverse impact of setting
a larger $x_1$ on the start-up delay can be compensated by the gain in the reduction of starvation probability.
Third, with the same sensitivity $\gamma$, the optimal $x_1^*$ of a long video stream can be smaller
than that of a short one in some situations. This is caused by the fact that the large threshold
might not significantly improve the starvation probability for a file of large size. We further
show the starvation probability in figure \ref{fig:fluid_qoe_starvprob}. A larger mean file size, or a
smaller $\gamma$ result in a larger probability of starvation. Unlike the start-up threshold $x_1^*$,
the starvation probability is shown to be strictly decreasing as $\lambda$ increases. 

\section{Conclusion and Discussion}
\label{sec:conclusion}

We have conducted an \emph{exact} analysis of the starvation
behavior in Markovian queues with a finite number of packet arrivals.
We perform a packet level analysis and a fluid level analysis. The packet level study
is carried out via two approaches, the Ballot theorem and the recursive equations. 
The former provides an explicit
expression, but is usually limited to i.i.d. packet arrivals. The latter can handle bursty 
packet arrival process, but without an explicit result.
From the perspective of a media service provider, 
we perform a fluid level analysis that computes the probability of starvation for a large number of video files. We further apply the theoretical results to tune the prefetching 
thresholds in order to optimize the objective QoE for media streaming services.

\begin{table}[!htb]
\centering
\begin{tabular}{|c|c|}
\hline {\bf Notation} & {\bf Definitions} \\
\hline& {\bf Section \ref{sec:Ballot}} \\
$\lambda$ & Packet arrival rate\\
\hline $\mu$ &  Packet service rate\\
\hline $\rho$ &   $\lambda/\mu$\\
\hline $p,\;q$ & $p=\lambda/(\lambda+\mu)$, $q=\mu/(\lambda+\mu)$ \\
\hline $x_1$ & Start-up threshold in pkts\\
\hline $T_1$ & Start-up delay\\
\hline $N$ & File size in pkts\\
\hline $P_s$ & Probability of starvation\\
\hline $P_s(j)$ & Probability of meeting $j$ starvations\\
\hline $\mathcal{E}(k_1)$ & First empty buffer after the service of $k_1$ pkts  \\
\hline {\scriptsize$\mathcal{S}_l(k_l,k_{l+1})$} & Empty buffer after the service of pkt $k_{l+1}$ \\
& given that the previous emptiness happens \\
& at the departure of pkt $k_{l}$\\
\hline $\mathcal{U}_j(k_j)$ & Last empty buffer observed after \\ 
& departure of packet $k_j$ \\
\hline $P_{\mathcal{E}(k_1)}$ & Probability of $\mathcal{E}(k_1)$ \\ 
\hline $J$ & Maximum number of starvations, $J=\lfloor\frac{N}{x_1}\rfloor$\\
\hline $d$ & Duration of a service slot in M/D/1 queue\\
\hline & {\bf Section \ref{sec:recursive}}\\
 $P_i(n)$ & Starvation probability of a file with $n$ packets, \\
& given that there are $i$ packets in the buffer\\
\hline $Q_i(k)$ & Probability that $k$ pkts out of $i$ \\
& leave the system during an inter-arrival period\\
\hline $P_{i}(j,n)$ & Probability of $j$ starvations with a file of size $n$, \\
& given that the first pkt sees $i$ pkts already there\\
\hline $\alpha$ & Transition rate from ON to OFF\\
\hline $\beta$ & Transition rate from OFF to ON\\
\hline $Q_i(k)^{ON}$ & Probability that $k$ pkts out of $i$ leave\\
& the system during an inter-arrival period at ON state\\
\hline& {\bf Section  \ref{sec:fluid}}\\
 $N_p$ & Total number of packets that are served\\
\hline $1/\theta$ & Mean of exponential file size distribution\\
\hline $N_m$ & Minimum file size for Pareto distribution\\
\hline $\upsilon$ & Exponent for Pareto distribution\\
\hline $\varrho$ & Mean of Normal distribtion for log-normal\\
\hline $\sigma$ & Standard deviation of Normal distribtion for log-normal\\
\hline&  {\bf Section \ref{sec:QoE}}\\
 $C_1(x_1)$ & QoE cost function for general situations\\
\hline $C_2(x_1)$ & QoE cost function for infinite media size with $\rho<1$\\
\hline {\scriptsize$LambertW$} & Lambert W-function\\
\hline
\end{tabular}
\caption{\small{Glossary of main notation}}
\label{table:variables1}
\end{table}

\section*{Appendix}

\subsection{Asymptotic Analysis}
We begin the asymptotic analysis with the following lemma.

\begin{lemma}
\label{lemma:no1}
Define a function $y(t) = \frac{1}{\sqrt{\pi t^3}} \exp\big( -\frac{v_1x^2}{t} - v_2t\big) $
where the constants $v_1$, $v_3$ and $x$ satisfy $v_2\geq 0, x\gg 0$ and $v_1\gg v_2$.
The integral $\int_x^{\infty}y(t)dt$ is approximated by
\begin{eqnarray}
\int_x^{\infty}y(t)dt \approx \frac{\exp\big( -2x\sqrt{v_1v_2}   \big)}{x\sqrt{v_1}}
\label{eq:lemma1}
\end{eqnarray}
\noindent with a degree of error $O(e^{-x})$.
\end{lemma}
\noindent \textbf{Proof:} We first show that $y(t)$ is a bounded function in
the range $t\in(0,\infty)$.
\begin{eqnarray}
\lim_{t\rightarrow 0}y(t) &{=}& \lim_{t\rightarrow 0}\frac{1}{\sqrt{\pi t^3}} \exp\big( -\frac{v_1x^2}{t}\big)  \nonumber\\
&{=}& \lim_{t\rightarrow 0}\frac{\exp\big( -\frac{v_1x^2}{t}\big)\cdot     \frac{v_1x^2}{t^2}}{\frac{3}{2} \sqrt{\pi t}} \approx \frac{2v_1x^2}{3t} \lim_{t\rightarrow 0}y(t) \nonumber
\end{eqnarray}
The above equation yields
\begin{eqnarray}
\lim_{t\rightarrow 0}y(t)\cdot\big(\frac{2v_1x^2}{3t}-1\big) = 0. \nonumber
\end{eqnarray}
Since the expression $\frac{2v_1x^2}{3t}+1$ approaches infinity as $t\rightarrow 0$,
there must exist $\lim_{t\rightarrow 0}y(t) = 0$. When $t$ increases to $\infty$, it
is easy to show $\lim_{t\rightarrow \infty}y(t) = 0$. Given that $y(t)$ is continuous
in $(0,\infty)$, it is also a bounded function.

Here, we suppose $v_2>0$. By differentiating $y(t)$ over $t$, we obtain
\begin{eqnarray}
\frac{dy(t)}{dt} = -\frac{\exp\big( -\frac{v_1x^2}{t}-v_2t\big)}{t^3 \sqrt{\pi t}}\big(v_2t^2 + \frac{3}{2}t - v_1x^2\big).
\end{eqnarray}
\noindent Letting the derivative $\frac{dy(t)}{dt}$ be 0, we obtain the optimal $t^*$ ($t^*>0$) to maximize $y(t)$, that is,
\begin{eqnarray}
t^* = \frac{\sqrt{9+16v_1v_2x^2} - 3}{4v_2}
\end{eqnarray}
When $t\leq t^*$, $y(t)$ is strictly increasing, and vice versa. The optimal value $t*$
is greater than $x$ if
\begin{eqnarray}
(v_1-v_2)\cdot x  \geq 3/2.
\label{eq:appendix_cond1}
\end{eqnarray}
Given that $v_1\gg v_2$ and $x$ is large, eq.\eqref{eq:appendix_cond1} is satisfied.
Therefore, the definite integral satisfies
\begin{eqnarray}
\int_0^{x} y(t)dt \leq x\cdot y(x).
\label{eq:appendix_cond2}
\end{eqnarray}
According to \cite{Book:MathFunc}(P1026, Chapter 29), the function $\frac{k}{2\sqrt{\pi t^3}}\exp{(-\frac{k^2}{4t})}$
has a Laplace transform $\exp(-k\sqrt{s})$. Therefore, one can easily
obtain the Laplace transform of $y(t)$ by
\begin{eqnarray}
y^*(s) = E[e^{-st}y(t)] =\frac{1}{x\sqrt{v_1}} \exp\big( -2x\sqrt{v_1(s+v_2)}   \big).
\label{eq:laplace1}
\end{eqnarray}
\noindent The integral $\int_0^{\infty} y(t)dt$ is obtained by
\begin{eqnarray}
\int_0^{\infty} y(t)dt = \lim_{s\rightarrow 0} y^*(s) = \frac{\exp\big( -2x\sqrt{v_1v_2}   \big)}{x\sqrt{v_1}}.
\label{eq:appendix_cond3}
\end{eqnarray}
\noindent The definite integral $\int_x^{\infty} y(t)dt$ satisfies
\begin{eqnarray}
\!\!\!\!\!\int_x^{\infty} y(t)dt \!\!\!&\geq&\!\!\!  \frac{\exp\big( -2x\sqrt{v_1v_2}   \big)}{x\sqrt{v_1}} - y(x)\cdot x \nonumber\\
\!\!\!&=&\!\!\! \frac{\exp\big( -2x\sqrt{v_1v_2}   \big)}{x\sqrt{v_1}} {-} \frac{\exp({-}v_1x{-}v_2x)}{\sqrt{\pi x}}.
\label{eq:appendix_cond4}
\end{eqnarray}
\noindent We hereby compar the two expressions $y_1 :=\frac{\exp\big( -2x\sqrt{v_1v_2}   \big)}{x\sqrt{v_1}}$
and $y_2 :=\frac{\exp({-}v_1x{-}v_2x)}{\sqrt{\pi x}}$ in eq.\eqref{eq:appendix_cond4},
\begin{eqnarray}
\frac{y_2}{y_1} = \sqrt{\frac{x}{\pi}} \cdot \exp\big(-x(\sqrt{v_1}-\sqrt{v_2})^2\big) .
\label{eq:appendix_cond5}
\end{eqnarray}
Given the conditions $v_1\gg v_2$, and $x\gg 1$, the ratio has $\frac{y_2}{y_1}\ll1$.
We can approximate the integral $\int_0^{\infty} y(t)dt $ by
\begin{eqnarray}
\int_0^{\infty} y(t)dt \approx \frac{\exp\big( -2x\sqrt{v_1v_2}   \big)}{x\sqrt{v_1}}.
\label{eq:approx1}
\end{eqnarray}
with a degree of error $O(e^{-x})$.

Next we consider a special case with $v_2=0$. Then $y(t)$ is rewritten as $y(t) = \frac{1}{\sqrt{\pi t^3}}
\exp\big(\frac{-v_1x^2}{t}  \big)$. It is easy to show that $y(t)$ is strictly increasing when
$t \leq \frac{2}{3}v_1x^2$, and strictly decreasing when $t > \frac{2}{3}v_1x^2$. Repeating
the above steps in the case $v_2>0$, we find
\begin{eqnarray}
\int_0^{\infty} y(t)dt \approx \frac{1}{x\sqrt{v_1}}
\label{eq:approx1}
\end{eqnarray}
\noindent which also matches eq.\eqref{eq:lemma1}. \done

\noindent \textbf{Approximating the starvation probability $P_s$:}

The starvation probability is given by
\begin{eqnarray}
P_{s} = \sum_{k=x_1}^{N-1} \frac{x_1}{2k-x_1}\binom{2k-x_1}{k}p^{k-x_1}(1-p)^k.
\end{eqnarray}
This equation contains the term obtained from the binomial distribution, which
is difficult to solve directly. We notice that the number of events $2k-x_1$
is usually large in term of the number of packets.
The variables $p$ and $q$ are not very much different
(otherwise the server is either over-provisioning or under-provisioning seriously).
This characteristic facilitates us to approximate the pdf of binomial distribution
by that of Gaussian distribution on the basis of central limit theory.
According to \cite{Book:BHH}, this approximation
is accurate enough if $(2k-x_1)p$ and $(2k-x_1)q$ are both greater than 5, or
the following inequality holds,
$\Big|\frac{1}{\sqrt{2k-x_1}}\big(\sqrt{\frac{q}{p}} - \sqrt{\frac{p}{q}}\big)\Big| < 0.3.$
The mean and the variance of the binomial distribution are $(2k-x_1)p$
and $(2k-x_1)pq$ respectively. Thus, there exists
\begin{eqnarray}
&&\binom{2k-x_1}{k-x_1}p^{k-x_1}(1-p)^k \;\;\; \thicksim \nonumber\\
&&\frac{1}{\sqrt{2\pi pq (2k-x_1)}}
\exp\big(-\frac{((k-x_1)-(2k-x_1)p)^2}{2pq(2k-x_1)}\big). \nonumber
\end{eqnarray}
\noindent To be more exact, the absolute error of \emph{c.d.f.} (integral of p.d.f), given by
the Berry-Ess��en theorem, is bounded by $0.7655(p^2 + q^2)/\sqrt{(2k-x_1)pq}$.
Then, the starvation probability is expressed as $P_{s}$
\begin{eqnarray}
\!\!\!\!\!\!\!\!\!\! &{\approx}&\!\!\!\!\! \sum_{k=x_1}^{\infty} \frac{x_1/(2k-x_1)}{\sqrt{2\pi pq (2k-x_1)}}
\exp\big(-\frac{((k-x_1)-(2k-x_1)p)^2}{2pq(2k-x_1)}\big) \nonumber\\
\!\!\!\!\!\!\!\!\!\!&=&\!\!\!\!\!  \sum_{k=x_1}^{\infty} \frac{x_1}{\sqrt{2\pi pq (2k-x_1)^3}}
\exp\big(-\frac{((2k-x_1)(\frac{1}{2}-p)-\frac{x_1}{2})^2}{2pq(2k-x_1)}\big) \nonumber\\
\!\!\!\!\!\!\!\!\!\!&{\approx}&\!\!\!\!\! \int_{x_1}^\infty  \frac{x_1\exp\big(-\frac{((2k-x_1)(\frac{1}{2}-p)-\frac{x_1}{2})^2}{2pq(2k-x_1)}\big)}{\sqrt{2\pi pq (2k-x_1)^3}}dk \label{eq:gaussianapprox2}\\
\!\!\!\!\!&{=}&\!\!\!\!\! \frac{x_1}{2\sqrt{2pq}}\exp{(\frac{(1-2p)x_1}{4pq})}\times \nonumber\\
\!\!\!\!\!&&\!\!\!\!\!
\int_{x_1}^\infty\frac{1}{\sqrt{\pi \widehat{k}^3}}\exp\big(-\frac{\widehat{k}(1-2p)^2}{8pq} -  \frac{x_1^2}{8pq \widehat{k}}\big) d\widehat{k} \\
\label{eq:gaussianapprox3}
\!\!\!\!\!&{\approx}&\!\!\!\!\! \frac{x_1}{2\sqrt{2pq}}\exp{(\frac{(1-2p)x_1}{4pq})}\times \nonumber\\
\!\!\!\!\!&&\!\!\!\!\!
\int_{0}^\infty\frac{1}{\sqrt{\pi \widehat{k}^3}}\exp\big(-\frac{\widehat{k}(1-2p)^2}{8pq} -  \frac{x_1^2}{8pq \widehat{k}}\big) d\widehat{k}.
\label{eq:gaussianapprox4}
\end{eqnarray}
\noindent The approximation in \eqref{eq:gaussianapprox2} is on the basis of the Riemann sum.
The approximation can be tightly bounded because the function to be integrated decreases to 0 exponentially.
The exact error bound can be obtained by computing both the right and the left Riemann sums.
The equality \eqref{eq:gaussianapprox3} is obtained by replacing $\widehat{k} = 2k-x_1$.
We replace $v_1$ and $v_2$ by $\frac{1}{8pq}$ and $\frac{(1-2p)^2}{8pq}$ respectively.
There has $\frac{v_1}{v_2} = \frac{1}{(1-2p)^2}$, which is much larger than 1 since
in realistic media streaming $p$ is not very far away from 1/2. The threshold $x$ is usually
large (i.e. more than 40 for the start-up delay of about 1s). Therefore,
the approximation in \eqref{eq:gaussianapprox4} from Lemma \ref{lemma:no1} is very tight.
Substituting $v_1$ and $v_2$ by the corresponding values in Lemma \ref{lemma:no1}, we derive
the asymptotic starvation probability as
\begin{eqnarray}
P_{s} &\approx& \exp\Big[\frac{x_1}{2pq}\big(\frac{1}{2}-p- |\frac{1}{2}-p|\big) \Big]
\label{eq:approx_prob_starv}
\end{eqnarray}
as the file size is large enough.
We next discuss the cases i)$\rho \leq 1$ and ii) $\rho>1$. If $\rho \leq 1$, or equivalently $p\leq\frac{1}{2}$,
eq.\eqref{eq:approx_prob_starv} is 1. If $\rho > 1$, or equivalently $p> \frac{1}{2}$, eq.\eqref{eq:approx_prob_starv}
is simplified as
\begin{eqnarray}
P_{s} \approx \exp\Big(\frac{x_1(1-2p)}{2pq}\Big).
\end{eqnarray}

\end{document}